\documentclass[pre,twocolumn,superscriptaddress,floatfix,aps,reprint,nofootinbib]{revtex4-2}
\usepackage{graphicx}
\usepackage{xcolor}
\usepackage{amsmath,amssymb,mathtools}   %
\usepackage{algorithm}
\usepackage{float}
\usepackage{algpseudocode}

\makeatletter
\newcommand{\parState}[1]{\State\parbox[t]{\dimexpr\linewidth-\ALG@thistlm}{#1\strut}}
\makeatother
\usepackage[pagebackref=false,breaklinks=true,colorlinks]{hyperref}
\usepackage{csquotes}
\hypersetup{colorlinks,breaklinks,
            urlcolor=[rgb]{0.803922, 0.0627451, 0.462745},
            citecolor=[rgb]{0.23137,0.4588,0.68627},
            linkcolor=[rgb]{0.803922, 0.0627451, 0.462745}
}

\renewcommand*{\selectlanguage}[1]{}

\begin{document}  

\title{Dynamics-preserving graph sparsification}

\author{Emmie Fitz-Gibbon}
\email{emmie\_fitz-gibbon@brown.edu}
\affiliation{Computational Neuroscience, Brown University, Providence RI, United States}
\affiliation{Santa Fe Institute, Santa Fe NM, United States}
\author{Maximilian Jerdee}
\thanks{Authors contributed equally.}
\affiliation{Santa Fe Institute, Santa Fe NM, United States}
\author{Yuanzhao Zhang}
\thanks{Authors contributed equally.}
\affiliation{Physics and Astronomy, University of Rochester, Rochester NY, United States}
\affiliation{Santa Fe Institute, Santa Fe NM, United States}
\author{Rosemary Braun}
\affiliation{Molecular Biosciences / Engineering Sciences \& Applied Mathematics / Physics \& Astronomy, Northwestern University, Evanston IL, United States}
\affiliation{Santa Fe Institute, Santa Fe NM, United States}
\author{Cristopher Moore}
\affiliation{Santa Fe Institute, Santa Fe NM, United States}

\begin{abstract}
Many dynamical systems unfold on networks of nodes coupled by edge interactions. This work is motivated by the following question: can we remove most edges from a networked dynamical system without significantly altering its behavior? For a given network, our goal is to find a sparser network that approximately preserves the dynamics on the nodes. We first study linear diffusion, where node values evolve according to the graph Laplacian. The dominant network sparsification method, effective resistance sampling, preserves all Laplacian eigenvalues within a uniform multiplicative bound. We show that with a limited edge budget, this is dynamically inefficient: diffusion is dominated by the slowest-decaying modes, so enforcing a uniform guarantee spends edges on fast modes that barely affect long-term dynamics. Based on this observation, we derive Diffusion Spectral Error (DSE), which evaluates in closed form the typical error in trajectories between the original and sparsified graphs over random initial conditions. Minimizing this error, which naturally prioritizes slow modes, yields sparse networks explicitly designed to preserve dynamics. At the same sparsification levels, DSE sparsification often preserves diffusion dynamics with several-fold lower error than effective resistance sampling, and up to two orders of magnitude lower on strongly modular networks. Although derived in a linear context, we empirically find that DSE sparsification gives similar improvements in the Kuramoto model (a nonlinear system of coupled oscillators), where DSE sparsification approximately preserves the synchronization threshold and, at strong coupling, order-parameter trajectories.

\end{abstract}

\maketitle

\section{Introduction}

Many physical, biological, and engineered systems evolve as dynamical processes on networks. Signals travel over neural synapses~\cite{bullmore_complex_2009}, infections spread between contacts~\cite{pastor-satorras_epidemic_2015}, and power grids synchronize over transmission lines~\cite{dorfler_synchronization_2010}. From a network of local interactions, these systems exhibit complex coordinated function and dynamics.

In this work we probe which properties of a network’s structure shape its dynamics. We ask which edges are required for its behavior and which are redundant, and seek a backbone of critical edges that approximately reproduce trajectories on the full network. By \emph{sparsifying} a network while retaining its dynamics, we can learn which structures its behavior depends on. Identifying these operative structures may guide the design of more efficient physical networks such as neuromorphic chips \cite{merolla_million_2014, davies_loihi_2018, benjamin_neurogrid_2014} or oscillator-based computing hardware \cite{csaba_coupled_2020, frady_robust_2019}, where each physical connection takes up space and consumes power \cite{bullmore_economy_2012}. 

The critical edges in a system are not always its strongest interactions. For instance, a chance encounter between far-flung travelers can have a greater impact on global pandemic spread than frequent contacts within an isolated town. Weak interactions that bridge otherwise disconnected parts of a system have an outsized effect on its overall behavior~\cite{granovetter_strength_1973}. 

\begin{figure*}[t]
    \centering
    
    \includegraphics{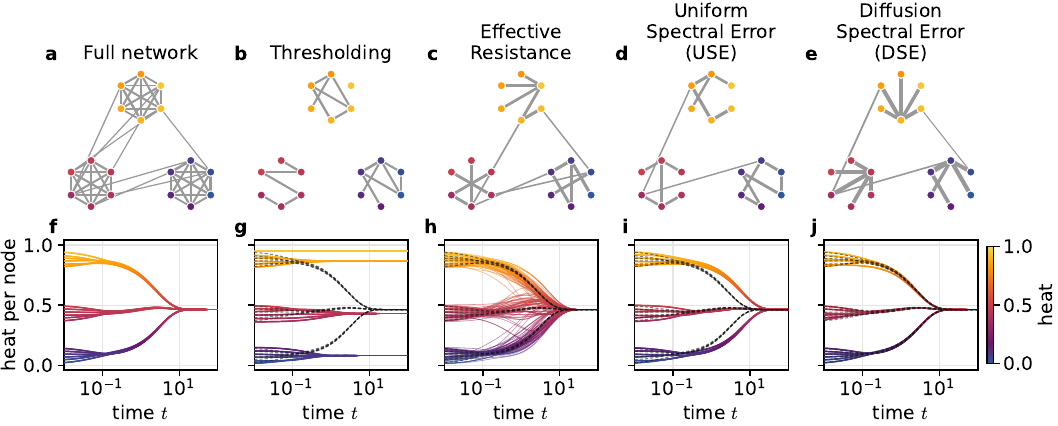}
    \caption{Linear diffusion can be thought of as heat spreading across a network, with conductance determined by edge weight. This example demonstrates how four sparsification approaches impact linear diffusion trajectories. \textbf{(a)} Example network of 18 nodes with three well-connected modules and a few weak bridging edges. Edge weights are represented by thickness, and node colors show starting heat. \textbf{(f)} The colored curves show how the heat on each node evolves over time, where each curve is colored by its current heat. Each community starts at a different temperature with minor variation between nodes. We observe initial synchronization within modules followed by convergence of the modules with each other. \textbf{(b)} Sparsification by weight thresholding removes all low weight edges, which in this example, disconnects modules \textbf{(g)} that then fail to synchronize. \textbf{(c)} Sparsification by the effective resistance sampling algorithm, and \textbf{(h)} trajectories plotted over 10 random network realizations (as the sampler is stochastic). \textbf{(d)} Uniform Spectral Error sparsification, which directly minimizes the error metric corresponding to effective resistance, \textbf{(i)} produces a network that equilibrates slower than the original.  \textbf{(e)} Diffusion Spectral Error sparsification, \textbf{(j)} where the diffusion trajectory closely matches that of the full network, especially at late time~$t$.}
    \label{fig:fig1}

\end{figure*}

Yet in many applied settings, weak interactions are often treated as unimportant. In dense empirical graphs such as functional brain connectomes~\cite{garrison_stability_2015} and gene co-expression networks~\cite{zhang_general_2005}, edges whose weights fall below a threshold are routinely discarded as a data cleaning step. While this suppresses noise, it can also remove the bridges that hold a system together. Figure~\ref{fig:fig1} illustrates this problem with a modular network -- weight thresholding fragments the network into separate communities on which heat diffusion no longer reaches a common temperature.

More principled sparsification approaches originated from theoretical computer science, beginning with cut sparsifiers, which preserve the total weight crossing every cut of the graph to within a multiplicative factor \cite{benczur_approximating_1996}. Spielman and Teng strengthened this notion with spectral sparsifiers, which preserve the Laplacian quadratic form, and therefore every Laplacian eigenvalue to within a multiplicative factor \cite{spielman_spectral_2011}. Spielman and Srivastava then showed that spectral sparsifiers can be constructed in nearly linear time by sampling edges in proportion to their \textit{effective resistance}, which is large for edges with few alternative paths between their endpoints, and naturally favors bridging edges \cite{spielman_graph_2011}. Further background on spectral sparsifiers and effective resistance sampling is detailed in Section \ref{sec:sparsifiers}.

Spectral sparsifiers bound the error in every eigenvalue by the same factor, treating all eigenmodes as equally important. A uniform spectral guarantee, however, can be poorly matched to the goal of preserving dynamics, because dynamical systems on networks are rarely equally sensitive to all eigenmodes. In linear diffusion, for example, the fastest modes decay almost immediately, while the slowest modes govern the long-term relaxation toward equilibrium. With a limited edge budget, spending edges to preserve the fast eigenvalues accurately comes at the expense of the slow modes that dominate the long-term dynamics (Figure \ref{fig:fig1}). Additionally, in many applications, the vertices of a graph represent specific, identifiable entities: genes, neurons, or locations in a transportation network. In these settings, it may not be sufficient to preserve aggregate spectral properties of the graph; we may also wish to preserve the spatial organization of its low-frequency modes. For example, two graphs with similar Laplacian spectra need not exhibit similar diffusion between particular \textit{pairs} of vertices. This motivates a desire for sparsifiers that preserve not only the spectrum but also the eigenvectors of the graph Laplacian. 

\paragraph{Contribution.}
In this paper we present a sparsification framework designed to preserve network dynamics, with a particular focus on diffusion dynamics. We introduce \emph{Diffusion Spectral Error} (DSE), an error metric that penalizes dynamically relevant network changes, and \emph{DSE sparsification}, an algorithm that minimizes DSE under a fixed edge budget. We define DSE as the expected error in linear diffusion trajectories between the original and sparsified networks: a quantity that is entirely determined by the Laplacian eigenmodes of each network, and disproportionately influenced by the slow modes. Thus, DSE sparsification prioritizes preserving the slow eigenvalues while \textit{also} maintaining the alignment of their corresponding eigenvectors.

DSE sparsification differs from effective resistance sampling in two ways: it optimizes edge selection directly rather than employing random sampling, and it prioritizes dynamically relevant modes rather than uniformly controlling the worst-case relative eigenvalue error. To separate these two effects, we also introduce and evaluate a spectral sparsifier achieved through optimization rather than sampling, which we call \textit{Uniform Spectral Error} (USE) sparsification, as described in Section \ref{sec:USE}.

\paragraph{Related Work.}
Prior work on sparsifying networked dynamical systems has largely applied effective resistance sampling directly. Although not tailored to preserve any particular dynamics, effective resistance sampling has been shown to preserve features of epidemic spreading~\cite{mercier_effective_2022}, point-vortex flows~\cite{nair_network-theoretic_2015}, and chimera states in Kuramoto--Sakaguchi oscillators~\cite{forrow_functional_2018}.

With the goal of preserving reaction diffusion dynamics \cite{ajayakumar_sgrdn-data_2026}, Ajayakumar and Raha used a subset of example trajectories and Laplacian constraints to \textit{learn} a sparse network. In this work, we avoid requiring simulated trajectories for our sparsification algorithm, for two reasons. First, for generality, we do not assume such example trajectories will be available. Second, ideally, the mechanism behind a sparsification algorithm reveals which graph properties are important for the dynamics, which learning can obscure. 

Recent work by Babecki et al.\  proposed sparsifiers that preserve exactly the first $k$ eigenpairs of the Laplacian~\cite{babecki_spectrahedral_2025}, motivated by the fact that low-frequency modes capture a graph’s global structure. Because the constraint is exact, however, the achievable sparsity shrinks rapidly as $k$ grows, and modes beyond $k$ are left unconstrained. DSE instead weights errors in every mode continuously by its dynamical relevance, allowing a smooth trade-off between accuracy and sparsity under a fixed edge budget.

\paragraph{Organization.}
The remainder of the paper is organized as follows. In Section~\ref{sec:sparsifiers}, we review spectral sparsifiers and describe the sparsification algorithms compared in our experiments. In Section \ref{sec:networks}, we detail the synthetic and empirical networks used for our simulations. In Section~\ref{sec:diffusion}, we analyze how each sparsifier distorts the Laplacian spectrum and demonstrate DSE's consistent advantage in preserving linear diffusion trajectories. In Section~\ref{sec:kuramoto}, we extend DSE to nonlinear dynamics via the Kuramoto model, showing that DSE preserves collective behavior such as the late-time order parameter. Section \ref{sec: discussion} reviews our central findings, and discusses how to extend our sparsification framework to a broader class of dynamical systems.

\section{Sparsifiers}
\label{sec:sparsifiers}
Consider an undirected, weighted graph $G=(V, E, w)$ with positive edge weights. The goal of graph sparsification is to approximate $G$ with a sparser graph $\widetilde{G} = (V, \widetilde{E}, \widetilde{w})$ on a subset of edges with potentially different weights, $\widetilde{E} \subset E$. If $\widetilde{G}$ is close to $G$ in some relevant metric, then $\widetilde{G}$ can be used in place of $G$ at lower memory and computation costs in simulations, or wiring costs in physical hardware. The choice of such a metric depends on the target application. Here, our goal is to find a sparse network that preserves the dynamical behavior of a dense network. However, much of the existing work on graph sparsification was motivated by the goal of speeding up the process of solving symmetric diagonally dominant linear systems. With this objective in mind, Spielman and Teng introduced the notion of spectral sparsifiers, which require that the Laplacian quadratic form of the sparse graph approximate that of the original \cite{spielman_spectral_2011}. Specifically, $\widetilde{G}$ is a $\varepsilon$-spectral sparsifier of $G$ if for all $x \in \mathbb{R}^n$ 
\begin{equation}
    (1 - \varepsilon)\, x^{T} L x
    \;\le\;
    x^{T} \widetilde{L}\, x
    \;\le\;
    (1 + \varepsilon)\, x^{T} L x
    \qquad 
    \label{eq:eff_r_bound}
\end{equation}

\noindent where $L = D - A$\footnote{$A$ is the weighted adjacency matrix and $D$ is the diagonal matrix of weighted degrees} is the graph Laplacian of $G$, and $\widetilde{L} = \widetilde{D} - \widetilde{A}$ is the graph Laplacian of $\widetilde{G}$. Here the Laplacian quadratic form is the total weighted disagreement across edges,
\begin{equation}
    x^{T} L x = \sum_{(i,j) \in E} w_{ij}\,(x_i - x_j)^2 ,
\end{equation}
and likewise for $\widetilde{L}$ over $\widetilde{E}$ with weights $\widetilde{w}$.
Intuitively, spectral sparsifiers require that for any assignment of values $x$ to the nodes, the total weighted 
disagreement across edges in $\widetilde{G}$ must be within a factor of $1 \pm \varepsilon$ of that in $G$. In particular, every eigenvalue of $\widetilde{L}$ lies within a factor $1 \pm \varepsilon$ of the corresponding eigenvalue of $L$, although the corresponding eigenvectors may differ.

\subsection{Effective Resistance Sampling}
Spielman and Srivastava \cite{spielman_graph_2011} showed that spectral sparsifiers 
can be efficiently constructed by \emph{effective resistance} sampling. Treating $G$ as an electrical network in which each edge is a resistor of 
conductance $w_{ij}$, the effective resistance between nodes $i$ and $j$ is
\begin{equation}
    R_{ij} = (e_i - e_j)^{\top} L^{+} (e_i - e_j) 
\end{equation}

where $L^{+}$ is the pseudoinverse of $L$.
The algorithm samples $q$ edges independently with replacement, where the probability $p_{ij}$ of sampling an edge $(i,j)$ is proportional to $w_{ij} R_{ij}$. The graph begins empty, and each sampled edge is added to $\widetilde{G}$ with weight $w_{ij}/(q\,p_{ij})$ rescaled by its probability of being sampled so the Laplacian is preserved in expectation, $\mathbb{E}[\widetilde{L}] = L$. With $q = O(n \log n / \varepsilon^2)$ samples, the sparsified network's~$\widetilde{L}$ concentrates to the original~$L$ within the bound from Eq.~\eqref{eq:eff_r_bound} and so controls its eigenvalues with probability at least $1/2$.

Effective resistance sampling runs in nearly linear time, and it naturally captures the ``strength of weak ties'' because bridging edges with few alternative paths carry high effective resistance and are therefore sampled with higher probability. 

At the same time, it has two notable drawbacks. First, the uniform spectral guarantee can be ill-suited to preserving \textit{dynamics} on a network, which can be non-uniformly impacted by different portions of the spectrum. Second, effective resistance sampling selects edges by random sampling, which satisfies the goal of nearly-linear runtime, but introduces run-to-run variability and can lead to suboptimal networks (including disconnected networks), particularly at low edge budgets. Indeed, while this sampling satisfies Eq.~\eqref{eq:eff_r_bound} in expectation, the probability of violating these bounds for a given realization generally increases as the edge budget decreases.

In the experiments to follow, we use independent effective resistance sampling as described in~\cite{spielman_sparsification_2015}, with the sampling rate calibrated to a prescribed edge budget. Specifically, each edge is kept independently with probability $p_{ij} = \min(1, a\,w_{ij}R_{ij})$, where $a$ is chosen so that the expected number of kept edges equals the budget. Sampled edges are assigned weights $w_{ij}/p_{ij}$ (i.e., rescaled by the sampling probability) ensuring that $\mathbb{E}[\tilde L] = L$. At the same number of edges, this independent inclusion sampling method preserves linear diffusion dynamics better than sampling with replacement, as shown in Appendix \ref{sec:er_sampling}.

\begin{figure*}[tp]
    \centering
    \includegraphics[width=1\linewidth]{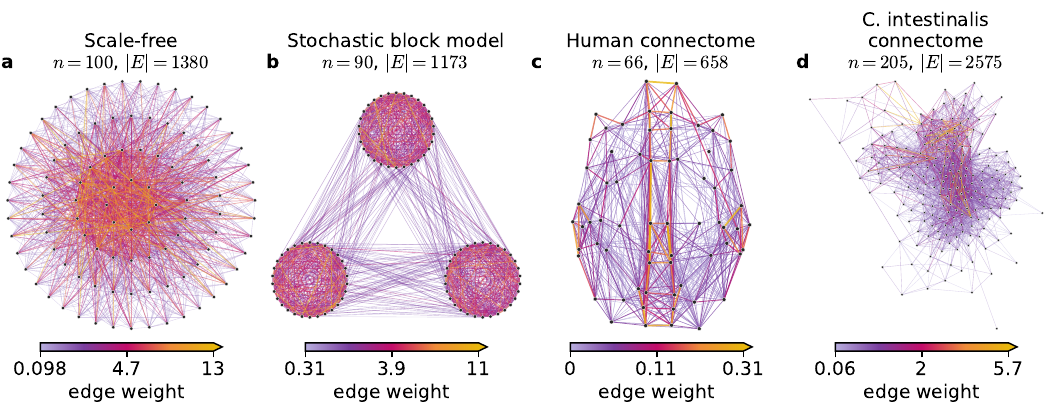}
    \caption{Four representative networks presented in the main text: \textbf{(a)} a weighted scale-free network, \textbf{(b)} a stochastic block model sample, \textbf{(c)} the human connectome, and \textbf{(d)} the \textit{C.~intestinalis} larval connectome. Line width and color represent edge weight according to each network's color scale. Further details are given in Appendix~\ref{sec:all_networks}. }
    \label{fig:four_networks}
\end{figure*}

\subsection{Diffusion Spectral Error sparsification}
\label{sec: DSE}

Instead of random sampling based on an importance score like effective resistance, we propose a sparsification scheme that directly minimizes an error metric that weights deviations in the Laplacian by their relevance to the dynamics. To find this error metric, we seek a measure of how well $\widetilde{G}$ reproduces the diffusion trajectories of $G$. A natural choice is the squared distance between trajectories from the same initial condition integrated over time. Linear diffusion on a network is governed by $\dot{x} = -Lx$, which admits the closed form solution $ x(t) = e^{-Lt}x_0$ from initial condition~$x_0$. We define trajectory error $\Delta(x_0)$ from a given initial condition as the squared distance between $x(t)$, the trajectory on the original network, and $\widetilde{x}(t)$, the trajectory on the sparsified network, integrated over time:  
\begin{equation}  
\label{eq:trajectory_error}
\Delta(x_0) = \int_{0}^{\infty} \|x(x_0,t) - \widetilde{x}(x_0, t)\|^{2} \,dt 
\end{equation}  
We average this trajectory error over isotropic, normally distributed random initial conditions to obtain $\Phi$, our Diffusion Spectral Error metric: 
\begin{align}
    \Phi \coloneqq \mathop{\mathbb{E}}_{x_0\sim\mathcal{N}(0,I)}[\Delta(x_0)].
\end{align}

As derived in Appendix~\ref{sec:deriving_dse}, the DSE may be written in terms of only the eigenvectors and eigenvalues~$(v_i,\lambda_i)$ of~$L$ and those~$(\widetilde{v}_j,\widetilde{\lambda}_j)$ of~$\widetilde{L}$ as
\begin{equation}
\Phi
 = \sum_{i,j \geq 2}
  \langle v_i,\tilde v_j\rangle^2
  \frac{(\lambda_i-\widetilde{\lambda}_j)^2}
       {2\,\lambda_i\,\widetilde{\lambda}_j\,(\lambda_i+\widetilde{\lambda}_j)}.
\label{eq:dse}
\end{equation}
This form lets us exactly evaluate the performance of any candidate~$\widetilde{G}$ directly from its adjacency matrix without the expensive simulation of many diffusion trajectories as we search for a sparse graph that minimizes~$\Phi$.

This form also reveals DSE's focus on slow modes compared to effective resistance sampling. Notice that for aligned modes,~$\langle v_i, \widetilde{v}_j \rangle \approx \delta_{ij}$, with small relative mismatch, $\widetilde{\lambda}_i = \lambda_i (1 + \eta)$ for~$\eta \ll 1$, each term reduces to $\eta^2/(4\lambda_i)$. The relative error in each mode is therefore weighted by $1/\lambda$, the mode's relaxation time.  This means that an optimization algorithm minimizing $\Phi$ will prioritize the slow modes with small~$\lambda$, as desired.

\paragraph*{Optimization.}
The optimization process (Algorithm \ref{alg:optimization} in Appendix \ref{sec:optimization_details}) takes the following shape. (1) We begin by taking a maximum spanning tree to ensure $\widetilde{G}$ is connected, with $\tilde{w}_{ij} = w_{ij}$. (2) Because the spanning tree may contain fewer edges than desired, we greedily add whichever edges most reduce $\Phi$ up to the edge budget, with $\tilde{w}_{ij} = w_{ij}$. (3) We then perform a series of greedy swaps of an included edge for an excluded edge if the exchange reduces $\Phi$. Specifically, we remove edge $(i,j)$ from $\widetilde{G}$ and add edge $(k,l)$ with $\tilde w_{kl} = (\tilde w_{ij}/w_{ij})\, w_{kl}$. (4) We then fix our choice of edges and optimize their weights with the numerical algorithm L-BFGS-B \cite{byrd_limited_1995}. Steps (3) and (4) are performed twice.  

\subsection{Uniform Spectral Error sparsification}
\label{sec:USE}
To disentangle the effects of direct optimization (rather than random sampling) from the effects of the DSE metric, we also directly optimize the error that Effective Resistance sampling controls: the $\varepsilon$ in Eq. (\ref{eq:eff_r_bound}).
The smallest $\varepsilon$ for which Eq. (\ref{eq:eff_r_bound}) holds is:
\begin{equation}
    \varepsilon(\widetilde{G}) \;=\; \max_{i \ge 2} |\mu_i - 1| \;=\; \bigl\| L^{+/2} \tilde L L^{+/2} - \Pi \bigr\|_2 ,
    \label{eq:seo_error}
\end{equation}
where $L^{+/2}$ denotes the square root of the pseudoinverse, $\mu_i$ are the nonzero-mode eigenvalues of $L^{+/2}\tilde L L^{+/2}$, $\Pi$ is the orthogonal projector onto the image of $L$, and $\|\cdot\|_2$ is the spectral norm. 

Uniform Spectral Error (USE) sparsification uses Algorithm \ref{alg:optimization}, the same optimization algorithm as DSE, with the error $\Phi$ replaced by $\varepsilon$. Because $\varepsilon$ is a maximum, it is not always differentiable as L-BFGS-B requires; We instead optimize the $p$-norm of the $|\mu_i - 1|$ with $p = 128$, a smooth surrogate for the maximum. (See Appendix~\ref{section: USE details} for further details). USE sparsification thus yields a spectral sparsifier through optimization, like DSE, rather than by random sampling.  Comparing USE with effective resistance sampling therefore isolates how much of DSE’s advantage comes from optimizing rather than from the metric.

\section {Networks}
\label{sec:networks}
To evaluate these three sparsification methods, effective resistance sampling, optimizing USE, and optimizing DSE, we measure how well they each preserve linear diffusion and Kuramoto dynamics on a range of networks. We focus on four representative networks (shown in Figure \ref{fig:four_networks}), consisting of two synthetic networks, and two biological networks. 

The synthetic networks are a weighted scale-free network generated with preferential attachment \cite{barabasi_emergence_1999} ($n = 100$, $d=15$ edges per new node), and a three-block stochastic block model \cite{holland_stochastic_1983} with dense, strongly connected blocks ($n = 90$, $p_{\mathrm{in}} = 0.7$, $p_{\mathrm{out}} = 0.085$). The biological networks are the coarse human connectome of Hagmann et al.\ \cite{hagmann_mapping_2008} ($n = 66$), and the \textit{C.~intestinalis} larval connectome \cite{ryan_cns_2016} ($n = 205$). For each of the experiments to come, the appendix includes results for a larger set of twelve networks. Appendix~\ref{sec:all_networks} details how each network was obtained or generated.

\section{Linear Diffusion Dynamics}
\label{sec:diffusion}
    
\begin{figure} [b]
    \centering
    \includegraphics[width=1\linewidth]{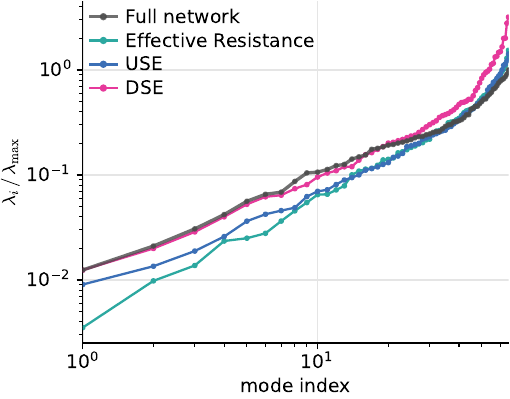}
    \caption{Laplacian eigenvalues of the human connectome ($n = 66$, $|E| =658$) and those of three sparsifications that keep 15\% of its edges. The eigenvalues are sorted from least to greatest, and normalized by the largest eigenvalue of the full graph. DSE closely matches the small eigenvalues (slow modes) and overshoots the large eigenvalues (fast modes), while USE and Effective Resistance overshoot the large eigenvalues less severely at the expense of undershooting the small eigenvalues.}
    \label{fig:connectome66_eigenvalues}
\end{figure}

\paragraph{DSE prioritizes slow modes.}
As we have argued, to preserve trajectories DSE prioritizes the slow modes of a network, which we now validate empirically. On the human connectome network, for example, DSE tracks the slow eigenvalues closely but \textit{inflates} the fast ones (Figure \ref{fig:connectome66_eigenvalues}). USE and Effective Resistance, by contrast, distort the spectrum more evenly across modes. This pattern holds across networks, as shown in Appendix Figure \ref{fig:eigenvalues_all_networks}.

\begin{figure*}[tp]
    \centering
    \includegraphics[width=1\linewidth]{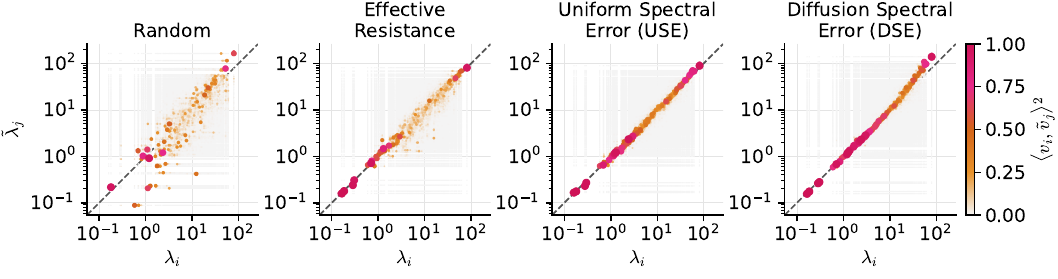}
    \caption{Eigenvector overlap between the C.~intestinalis connectome and four sparsifiers keeping 25\% of its edges. At every pair of modes between the original and sparse networks, ($\lambda_i$, $\widetilde{\lambda}_j$), we place a dot where the color and size show $\langle v_i, \tilde v_j\rangle^2$, how aligned the corresponding eigenvectors are. A perfect sparsifier would only have colored dots on the diagonal (every eigenvalue perfectly matched, and corresponding eigenvectors perfectly aligned).}
    \label{fig:eigenoverlap}
\end{figure*}

\paragraph{DSE maintains eigenvector overlap, particularly for slow modes.} Along with preserving the eigenvalues, we would also like the associated eigenvectors to remain well aligned as they represent the content of these modes. Figure \ref{fig:eigenoverlap} represents the overlaps~$\langle v_i, \widetilde{v}_j \rangle^2$ between each pair of eigenvalues by sized and colored dots located at their eigenvalue pair ($\lambda_i$, $\widetilde{\lambda}_j$), so that perfect alignment~$\langle v_i, \widetilde{v}_j \rangle = \delta_{ij}$ appears as a pink line along the diagonal. USE and DSE both show strong diagonal lines compared to the Effective Resistance and Random sampling. However, this plot also captures more benign changes, such as very similar eigenvalues with aligned eigenvectors, despite not having the same index after sorting ($i \neq j$). Or, in the case of DSE sparsification, overshoot the eigenvalues of the fastest modes, which is inconsequential for linear diffusion dynamics.

\paragraph{DSE preserves linear diffusion dynamics.}

\begin{figure}
    \centering
    \includegraphics[width=1\linewidth]{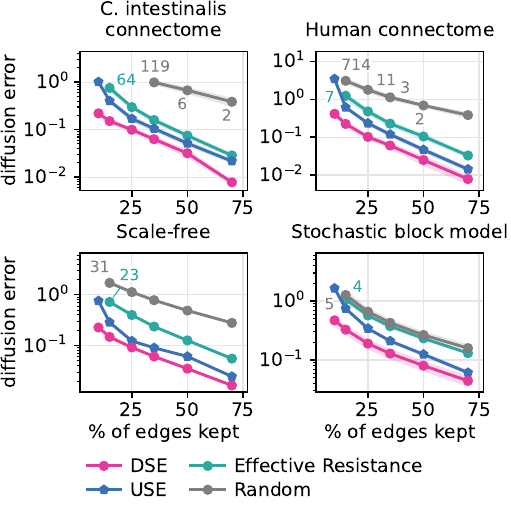}
    \caption{Diffusion Spectral Error (DSE) sparsification best preserves linear diffusion dynamics. Linear diffusion trajectory error simulated from 15 random initial conditions, with $x_{0}$ drawn from  $\mathcal N(0, I)$. Each point is the mean over 15 random initial conditions; for each initial condition, errors are first averaged over the 10 Effective Resistance or Random network draws and, for generated networks, over the 10 network realizations. Shading shows standard deviation, which is often narrower than the line itself. Printed numbers next to individual points give the mean number of re-draws necessary to yield a connected network.}
    \label{fig:linear_diffusion_error}
\end{figure}

Figure \ref{fig:linear_diffusion_error} shows linear diffusion trajectory error for DSE, USE, Effective Resistance sampling, and random sampling, which employs the same sampling algorithm as Effective Resistance, except the sampling probabilities are equal for each edge. Diffusion trajectory error is measured as the mean absolute difference between trajectories from the same initial condition, averaged over time, from times $t = 0$ to  $t = 3/\lambda_2$, where $\lambda_2$ is the smallest nonzero Laplacian eigenvalue from the full network. By $t = 3/\lambda_2$, the slowest mode has decayed to $e^{-3} \approx 5\%$ of its initial amplitude, so the window captures nearly the full relaxation without diluting the average with time spent at equilibrium. 

The time averaged error is then normalized by the full trajectory's average distance from equilibrium, $\langle |x(t) - \bar x| \rangle_{t}$, so the error is relative to how far the full network is from equilibrium over the simulated window: an error of 1 means the sparse trajectory is, on average, as far from the full trajectory as the full trajectory is from equilibrium, and 0.1 means it is a tenth as far. 

Since random sampling and Effective Resistance sampling are not deterministic, we draw ten sparse graphs per seed to mitigate the impacts of lucky or unlucky sampling draws. Effective Resistance and Random sampling can also produce disconnected graphs at low edge budgets, which would yield infinite $\Phi$, so we redraw (i.e. start over and run the sampling algorithm again) until we find a connected network. This ensures that the observed benefits of DSE are not simply due to the fact that it always produces a connected graph. The mean number of draws needed to yield a connected graph is printed beside each point, showing that, for example, Effective Resistance sampling on the \textit{C. intestinalis} connectome required 64 draws to yield a connected network. Note that conditioning on connectivity biases the samplers towards having more edges than the target, and for effective resistance, increases the probability that the retained realization satisfies the desired spectral bounds. For the two generated networks (Scale-free and Stochastic block model), we average over 10 network realizations. 

Figure \ref{fig:linear_diffusion_error} shows that DSE sparsification outperforms all other sparsifiers on all networks, and all levels of sparsity. Appendix Figure \ref{fig:diffusion_error_all} confirms that this holds across all twelve networks tested. Figure \ref{fig:linear_diffusion_error} also reveals that USE sparsification outperforms Effective Resistance Sampling, demonstrating that part of DSE's gain comes from optimization rather than sampling, but prioritizing the slow modes is a further improvement. That DSE performs best in these trials is, of course, unsurprising as it produces networks directly optimized for linear diffusion. The consistent several-fold improvement DSE realizes over alternative sparsification methods, however, gives insight into how much the other methods lose by spending edge budget on preserving dynamically irrelevant fast modes. 

\begin{figure} 
    \centering
    \includegraphics[width=\linewidth]{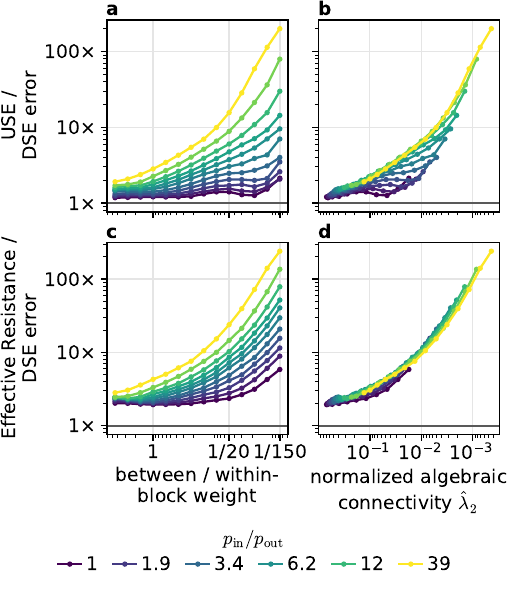}
    \caption{DSE's advantage over USE and Effective Resistance on linear diffusion trajectory error grows with bottleneck strength. We sweep over stochastic block models with 90 nodes and $p_{\mathrm{in}} = 0.35$, and sparsify each with DSE, USE, and Effective Resistance, keeping 25\% of the full network's edges. Each point is averaged over 40 realizations of the block model with the specified parameters, and Effective Resistance is averaged over 10 draws per realization, each redrawn until connected. The y-axis shows the linear diffusion error of USE (a, b) or Effective Resistance (c, d) divided by that of DSE, so that values $>1$ mean that DSE performed better. Linear diffusion error is measured as in Figure \ref{fig:linear_diffusion_error}. Each line is one value of $p_{\mathrm{in}}/p_{\mathrm{out}}$, the probability that two nodes in the same block are connected divided by the probability that two nodes in separate blocks are connected. \textbf{(a, c)} DSE's advantage over USE (a) and Effective Resistance (c) against the between/within-block weight ratio. The advantage is largest for networks with strong clustering (yellow) \textit{and} weak bridges between blocks, and small where edges are spread evenly between blocks, even when between-block edges are far lighter. \textbf{(b, d)} The same networks against bottleneck strength, measured by the normalized algebraic connectivity $\hat{\lambda}_2$~\cite{chung_spectral_1996}. Against Effective Resistance (d), all values of $p_{\mathrm{in}}/p_{\mathrm{out}}$ collapse onto a single curve.}
    \label{fig:sbm_sweep}
\end{figure}

 \paragraph{DSE's advantage is greatest on networks with strong bottlenecks.} Diffusion Spectral Error (DSE) sparsification preserves linear diffusion dynamics by prioritizing slow Laplacian eigenmodes over fast ones. DSE's advantage over Effective Resistance and Uniform Spectral Error (USE) is therefore greatest when there is a large gap between the slow and fast relaxation rates, as in strongly modular networks with extreme bottlenecks. This effect is most clearly demonstrated with Stochastic Block model networks. Figure \ref{fig:sbm_sweep}a demonstrates that in SBM networks, both strong clustering \textit{and} weak edges bridging communities are required for DSE to have a strong advantage over USE. With no clustering structure ($p_{\mathrm{in}} \approx p_{\mathrm{out}}$), an extreme between/within block edge weight ratio contributes much less to DSE's advantage.

A reasonable overall predictor of DSE's advantage is the normalized algebraic connectivity $\hat{\lambda}_2$, the second-smallest eigenvalue of $D^{-1/2} L D^{-1/2}$, which measures the strength of bottlenecks in a network. Through Cheeger's inequality, $\hat{\lambda_2}/2 \le h \le \sqrt{2\hat{\lambda_2}}$, it bounds the graph's conductance $h$ from above and below, so small $\hat{\lambda}_2$ indicates a strong bottleneck \cite{chung_spectral_1996}. Looking at DSE's advantage over Effective Resistance, $\hat{\lambda}_2$ is highly predictive. In Figure \ref{fig:sbm_sweep}(d), all $p_{\mathrm{in}}/p_{\mathrm{out}}$ ratios collapse onto a single curve. 
 
Normalized algebraic connectivity is also predictive of DSE's advantage over both USE and Effective Resistance in other network generator models (Appendix Figures \ref{fig:DSE_advantage_use} and \ref{fig:DSE_advantage_er}), but the effect is not as pronounced across all generator models as in SBM networks.

\section{Nonlinear Dynamics: The Kuramoto Model}
\label{sec:kuramoto}
\begin{figure}[t]
    \centering
    \includegraphics[width=1\linewidth]{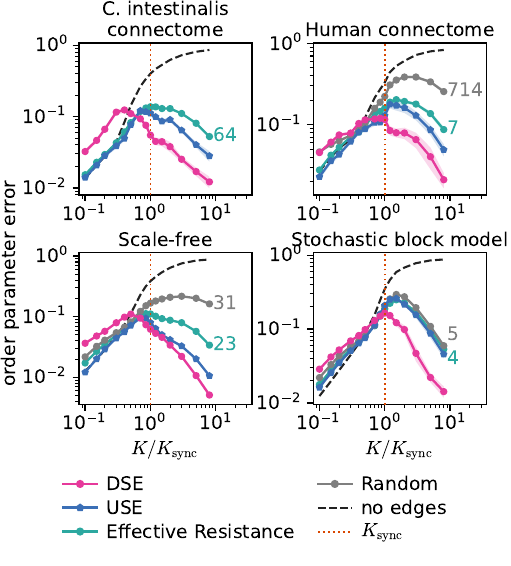}
    \caption{Kuramoto order-parameter error by coupling strength at 15\% of edges kept. The x-axis sweeps over  global coupling $K$ in units of each network's $K_{\mathrm{sync}}$, the coupling at which the full network's late-time order parameter reaches $r = 0.5$. The shading shows $\pm 1$ standard error of the mean over seeds.  Sampled networks are left out of a panel entirely if fewer than 1 in 1000 draws are connected. The number beside the right end of the Random and Effective Resistance curves gives the mean number of draws needed to yield a connected graph. The dashed black line shows the order parameter error with $K = 0$.}
    \label{fig:order_parameter_error}
\end{figure}

We have established that DSE preserves linear diffusion dynamics, but many physical and biological systems have \textit{nonlinear} couplings. In this section we explore graph sparsification for preserving the dynamics of the Kuramoto model, a nonlinear model of coupled oscillators \cite{araki_self-entrainment_1975}. Kuramoto dynamics are given by
\begin{equation}
    \dot\theta_i = \omega_i + K \sum_{j=1}^{n} w_{ij} \sin(\theta_j - \theta_i),
    \label{eq:kuramoto}
\end{equation}
where $\theta_i$ is the phase of oscillator $i$, $\omega_i$ is its natural frequency, $w_{ij}$ is the weight of the edge between nodes $i$ and $j$, and $K$ is the global coupling strength. Intuitively, each oscillator wants to run around the unit circle at its own natural frequency, but gets pulled toward the phases of its neighbors through the sine of their phase difference, in proportion to the connecting edge weights. The long term behavior of the Kuramoto model is governed by the underlying network, the spread of natural frequencies, and the global coupling constant K. With sufficiently strong coupling, the pull between oscillators overpowers the heterogeneity in the natural frequencies, and all oscillators will begin moving as a synchronized pack with a shared frequency. With moderate coupling, some fraction of oscillators will become frequency synchronized, while the others remain out of sync. The degree of synchronization in the Kuramoto model is measured by the order parameter
\begin{equation}
    r(t)\, e^{\mathrm{i}\psi(t)} = \frac{1}{n} \sum_{j=1}^{n} e^{\mathrm{i}\theta_j(t)},
    \label{eq:order_parameter}
\end{equation}
where $r = 1$ when all oscillators share the same phase and $r \approx 0$ when phases are spread uniformly.

Because the coupling between nodes is a nonlinear function of their difference, we cannot write down in closed form the expected trajectory error between $G$ and $\widetilde{G}$. However, when coupling is sufficiently strong, we can gain traction by looking at the dynamics near the synchronized state. Near phase synchronization, $\sin(\theta_j - \theta_i) \approx \theta_j - \theta_i$, and expressed in the frame rotating at the mean natural frequency $\bar\omega$, a perturbation $\delta = \theta - \theta^*$ around the synchronized state $\theta^*$ reduces to linear diffusion:
\begin{equation}
    \dot\delta = -K L \delta .
    \label{eq:kuramoto_linear}
\end{equation}
We therefore expect DSE to help preserve Kuramoto dynamics at strong coupling where the trajectory is dominated by diffusion-like dynamics. 

\paragraph{DSE preserves order parameter trajectories at high coupling.}
One way to compare behavior of Kuramoto dynamics on $G$ and $\widetilde{G}$ is to look at how the order parameter evolves in each network. We measure order parameter error as the time averaged difference in instantaneous order parameter: $\langle|r(t) - \tilde r(t)|\rangle_t$. 

Figure \ref{fig:order_parameter_error} shows the order parameter error for DSE, USE, Effective Resistance, and Random sampling by global coupling strength on the four representative networks. At each coupling the full and the sparse networks are integrated from the same natural frequencies and initial phases (Appendix \ref{sec:K_sync}). Each point is the mean over 15 seeds. Within a seed, Effective Resistance and Random are averaged over 10 draws and, for the generated networks, we average over 10 realizations. As in Figure \ref{fig:linear_diffusion_error}, the Effective Resistance and Random sparse networks are redrawn until connected, to rule out any disadvantage due to disconnection alone.

DSE preserves the order parameter trajectory better than all other methods past $K_{\mathrm{sync}}$, the coupling at which the full network's late-time order parameter reaches $r = 0.5$ (see Appendix \ref{sec:K_sync}). The dashed black line shows the order parameter error with all edges removed (i.e. $K = 0$), which shows that before $K = K_{\mathrm{sync}}$ the sparse graphs are often no better than simply removing all edges. Appendix Figure \ref{fig:order_parameter_all} confirms that this pattern holds across all networks except the Core-periphery network.

\paragraph{DSE preserves synchronization transition curves.}

Another way to quantify how well Kuramoto dynamics are preserved on $\widetilde{G}$ is to compare the phase transition curves between the original and sparse networks. The phase transition curve plots the late-time order parameter $r$ against coupling strength. In the infinite $n$ limit, this curve describes a phase transition where synchronization begins at zero before sharply increasing at a critical coupling value, and eventually plateauing. In finite networks, the curve typically takes a sigmoidal shape, which varies based on the underlying network. 

\begin{figure*} 
    \centering
    \includegraphics[width=1\linewidth]{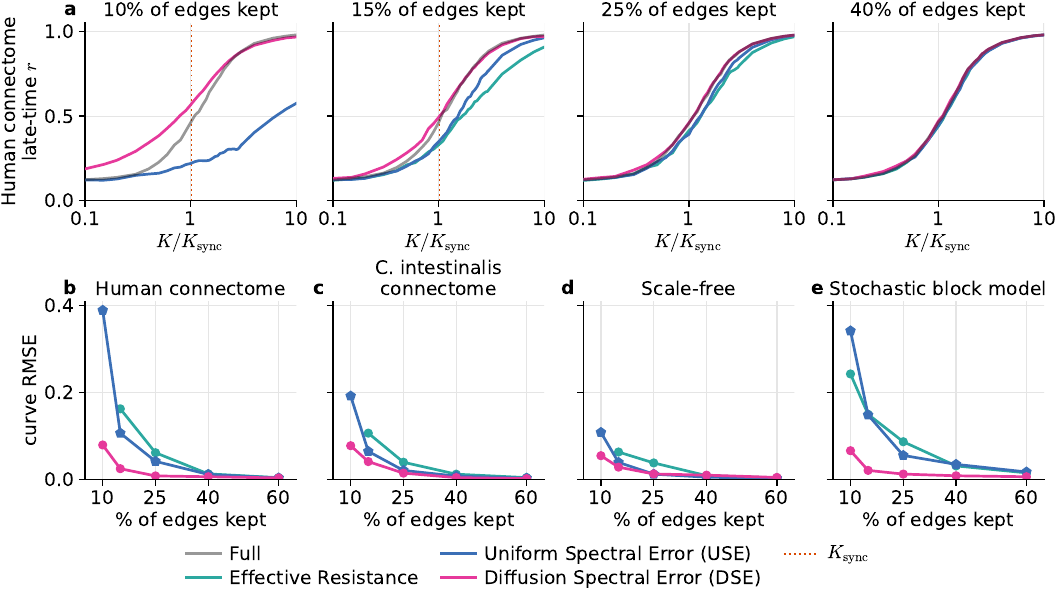}
    \caption{DSE best preserves the synchronization transition curve as global coupling is increased. \textbf{(a)} Late-time order parameter $r$ of the human connectome against $K/K_{\mathrm{sync}}$ at 10\%, 15\%, 25\% and 40\% of edges kept. Effective Resistance draws are redrawn until connected and left out where fewer than 1 in 1000 single draws connect, which happens for Effective Resistance in panel (a). \textbf{(b--e)} Root mean squared difference between each sparse curve and the full one, against the fraction of edges kept (10--60\%), for \textbf{(b)} the human connectome, \textbf{(c)} C.~intestinalis, \textbf{(d)} the scale-free network ($d =15$) and \textbf{(e)} the stochastic block model ($p_{\mathrm{out}} = 0.085$).}
    \label{fig:phase_transition_curve}
\end{figure*}
\begin{figure*} 
    \centering
    \includegraphics[width=1\linewidth]{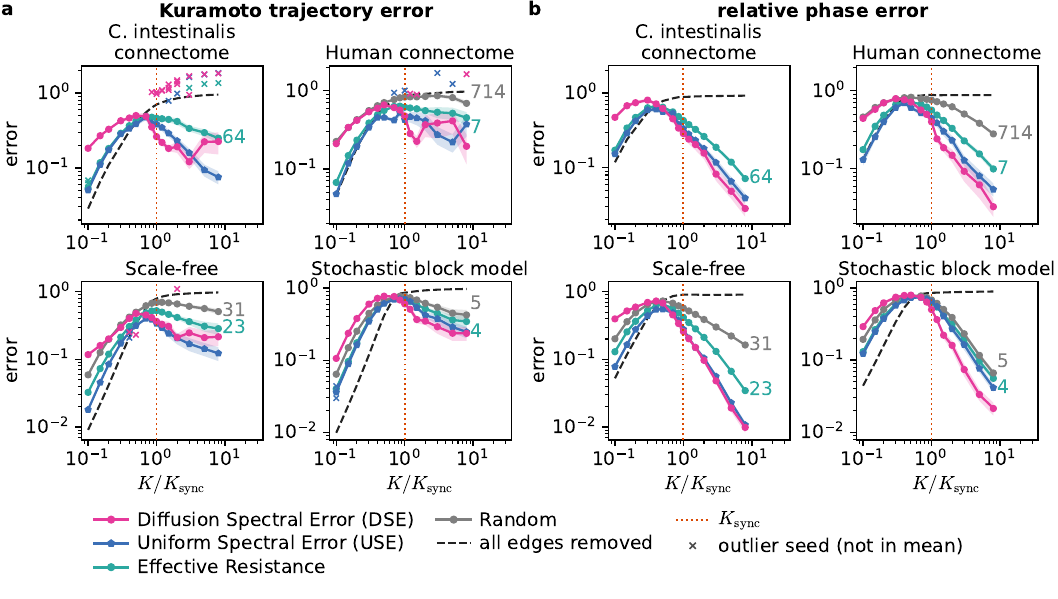}
    \caption{Kuramoto trajectory error by coupling strength at 15\% of edges kept. (a) Phase error, Eq. \eqref{eq:kuramoto_error}, and (b) relative phase error, measured after subtracting each run's mean phase from every oscillator, for the four representative networks. The x-axis sweeps over global coupling $K$ in units of each network's $K_{\mathrm{sync}}$, the coupling at which the full network's late-time order parameter reaches $r = 0.5$. The shading shows $\pm 1$ standard error of the mean over seeds; in (a), crosses mark outlier seeds, which are left out of the mean. Sampled networks are left out of a panel entirely if fewer than 1 in 1000 draws are connected. The number beside the right end of the Random and Effective Resistance curves gives the mean number of draws needed to yield a connected graph.}
    \label{fig:kuramoto_trajectory}
\end{figure*}

Figure \ref{fig:phase_transition_curve}a  shows the phase transition curves for the human connectome network and the DSE, USE, and Effective Resistance sparsified versions at various sparsity levels. The DSE curve, shown in pink, approaches the full curve far faster than the USE and Effective Resistance curves as more edges are included. Figure \ref{fig:phase_transition_curve}b--e shows the root mean squared error between phase transition curves for each of the four representative networks, where DSE shows the lowest error, particularly at 10\% to 15\% of edges kept. Appendix Figure \ref{fig:phase_transition_curves} shows that DSE has the lowest error at low edge budgets (10–15\%) on all networks except core–periphery, and at higher budgets USE and DSE are both close to the full curve. See Appendix \ref{sec:synch_curves} for further details on computing the phase transition curves. 

\paragraph{DSE preserves relative phases, but not the global phase.} Beyond collective behaviors like synchronization, we can also measure how well each sparsifier preserves the full $n$-dimensional Kuramoto trajectories. We measure Kuramoto trajectory error by the phase difference between corresponding oscillators,
\begin{equation}
    E_K = \bigl\langle 1 - \cos\bigl(\theta_i(t) - \tilde\theta_i(t)\bigr) \bigr\rangle,
    \label{eq:kuramoto_error}
\end{equation}
averaged over oscillators $i$ and times $t \in [0, 40]$, from the same initial phases and natural frequencies on both networks. $E_K = 0$ when the phases agree, $E_K \approx 1$ when they are uncorrelated, and $E_K = 2$ when they are opposite.
Figure 
\ref{fig:kuramoto_trajectory}a shows that although DSE and USE both outperform Effective Resistance and Random sampling at high coupling constants, DSE is not consistently better than USE at preserving the full, $n$-dimensional trajectory. Figure~\ref{fig:kuramoto_trajectory}b shows the relative phase error, which is the trajectory error measured after subtracting each run's mean phase $\psi(t)$ from every oscillator, $\langle 1 - \cos\bigl((\theta_i - \psi) - (\tilde\theta_i - \tilde\psi)\bigr) \rangle_{i,t}$, so that a uniform rotation of all phases does not count as error. In other words, relative phase error is the trajectory error in the rotating frame.

Figure~\ref{fig:kuramoto_trajectory} shows that DSE outperforms or ties with USE in relative phase error at high coupling, but shows no clear advantage for Kuramoto trajectory error. This implies that DSE struggles to preserve the overall phase of the pack, which is likely set during disordered initial transients, where the approximation of Eq. \eqref{eq:kuramoto_linear} does not hold. Appendix Figures~\ref{fig:relative_phase_all} and ~\ref{fig:kuramoto_trajectory_all} show relative phase error and Kuramoto trajectory error for all twelve networks. 

\section{Discussion}
\label{sec: discussion}

Any sparsifier with a fixed edge budget implicitly decides which characteristics of the graph to preserve or sacrifice. Our central argument is that uniform spectral preservation is not the optimal objective for preserving network dynamics. Instead, we directly minimize an error metric that can be adapted to prioritize the dynamically relevant portions of the spectrum. With linear diffusion, the importance of each mode can be analytically derived, yielding our DSE metric. We demonstrate that DSE sparsification protects small eigenvalues (associated with slow modes) and corresponding eigenvectors in a variety of network examples, and that this translates into a considerable improvement in preserving diffusion trajectories on the sparsified graphs. We demonstrated that this advantage grows as the normalized algebraic connectivity $\hat{\lambda}_2$ decreases, indicating increasingly pronounced network bottlenecks.

While we cannot derive an error quantity analogous to DSE for \textit{nonlinear} dynamical systems, we demonstrate that for the Kuramoto model -- a nonlinear system that behaves similarly to linear diffusion around the synchronous state -- DSE sparsification preserves collective behavior such as synchronization thresholds. 

\paragraph{Generalizations.}
Linear diffusion is a particular example of network dynamics with closed form solutions that can be expressed in the Laplacian eigenbasis, which allowed us to derive DSE. However, the same derivation can apply to other linear dynamical systems. Note that the only property of linear diffusion used to obtain the closed form expression for $\Phi$ in Section ~\ref{sec: DSE} is that the propagator $e^{-Lt}$ is diagonal in the eigenbasis of $L$. The same derivation therefore applies to any linear
dynamical system whose propagator is a function of the Laplacian, $x(t) = h(L,t)\,x_0$,
giving:

\begin{equation}
    \mathcal{E} = \sum_{i,j} \langle v_i, \widetilde{v}_j \rangle^2 \, f(\lambda_i, \widetilde{\lambda}_j),
    \label{eq:general_form}
\end{equation}
\begin{equation}
    f(\lambda, \widetilde{\lambda}) = \int_0^\infty \bigl[ h(\widetilde{\lambda}, t) - h(\lambda, t) \bigr]^2 \, dt .
\end{equation}

This suggests a more general interpretation of $\mathcal{E}$. Each eigenvalue mismatch is weighted by the overlap $\langle v_i, \widetilde{v}_j \rangle^2$, so deviations count only to the extent that the corresponding eigenvectors are aligned. The dynamics, through the propagator $h(\lambda, t)$, determine the cost of a mismatch in each part of the spectrum.

\paragraph{Limitations and Future Work.}
The main limitation of DSE sparsification is the time complexity of the optimization algorithm, which runs in polynomial time. A key advantage of effective resistance sparsification is the nearly linear runtime, which allows sparsification of extremely large networks. Conversely, DSE sparsification is most advantageous on relatively small networks, where random sampling produces high variance, and the computationally expensive optimization process is feasible. Reducing the time complexity of Algorithm ~\ref{alg:optimization} is a key goal for future work. 

Further, evaluating the performance of DSE sparsification for nonlinear network dynamics relies mostly on empirical evidence. Future work could explore the applicability of DSE sparsification to other nonlinear systems, such as reaction--diffusion systems.

Another interesting extension of this work would be graph sparsification to preserve dynamical systems that depend largely on the fastest modes, which, in the framework of Eq. \eqref{eq:general_form}, would simply involve replacing the function $f$ with one that prioritizes large eigenvalues rather than small ones. A similar strategy could be used to target mid-spectrum properties. 

Finally, the presented sparsification strategy applies only to undirected networks with positive edge weights, which, conveniently, yield symmetric graph Laplacians with non-negative eigenvalues. Extending these methods to networks that include negative edge weights is less straightforward. One potential approach is to integrate Eq. \eqref{eq:trajectory_error} up to a finite time $T$ to prevent blow-up, which would mean prioritizing whichever modes dominate the trajectory between times $t = 0$ and $t = T$.

\begin{acknowledgments}

The authors of this work are grateful to Leo Kozachkov and Alexander Mercier for helpful discussions, and the Santa Fe Institute Undergraduate Complexity Research program team for their support. This work was supported in part by the National Science Foundation under Award No. 2349052, ``REU Site: Quantitative Rules of Life: General Theories across Biological Systems'' and the Robert Wood Johnson Foundation.

\end{acknowledgments}

\bibliography{references}

\appendix

\section{Deriving Diffusion Spectral Error}
\label{sec:deriving_dse}
We begin with trajectory error $\Delta$ from a given initial condition $x_0$, which we define as the squared distance between $x(t)$, the trajectory on the original network, and $\widetilde{x}(t)$, the trajectory on the sparsified network, integrated over time:  

\begin{equation}  
\Delta(x_0) = \int_{0}^{\infty} \|x(x_0,t) - \widetilde{x}(x_0, t)\|^{2} \,dt 
\end{equation}  
We take the expected value over isotropic normally distributed initial conditions to obtain $\Phi$, our Diffusion Spectral Error metric. 
\begin{equation}
\Phi
\coloneqq
\mathop{\mathbb{E}}_{x_0\sim\mathcal{N}(0,I)}[\Delta(x_0)]
=
\int_{0}^{\infty}
\bigl\|e^{-\widetilde{L}t}-e^{-Lt}\bigr\|_F^2\,dt
\label{integral_error}
\end{equation}
Let $v_i$ be the eigenvectors of $L$ with associated eigenvalues $\lambda_i$, and $\widetilde{v}_j$ be the eigenvectors of $\widetilde{L}$ with associated eigenvalues $\widetilde{\lambda}_j$. Since $L$ and $\widetilde{L}$ are symmetric, $\{v_i\}$ and $\{\tilde v_j\}$ are orthonormal bases. Note that if $G$ and $\widetilde{G}$ are connected, both Laplacians share $\lambda_1 = \widetilde{\lambda}_1 = 0$ with
$v_1 = \widetilde{v}_1 = \mathbf{1}/\sqrt{n}$, so we can drop the zero modes from here on. As with any set of orthonormal bases, the Frobenius norm can be written as:  $\|A\|_F^2 = \sum_{i,j} (v_i^T A \widetilde{v}_j)^2 $.
Recall that $v_i^\top e^{-L t} = e^{-\lambda_it}v_i^\top $ and $e^{-\tilde L t}\,\tilde v_j = e^{-\widetilde{\lambda}_j t}\,\tilde v_j$, so we can rewrite $\Phi$ as a function of the two eigenbases.

\[ \Phi = \int_0^\infty
   \sum_{i,j \geq 2}
   \bigl(
     v_i^\top e^{-\widetilde{\lambda}_j t}\,\tilde v_j
     -
     v_i^\top e^{-\lambda_i t}\,\tilde v_j
   \bigr)^2\,dt 
\]
   \[  = \sum_{i,j \geq 2}
   \langle v_i,\tilde v_j\rangle^2
   \int_0^\infty
   \bigl(
     e^{-\widetilde{\lambda}_j t}
     -
     e^{-\lambda_i t}
   \bigr)^2\,dt 
\]

Expanding the squared term and evaluating the integral:
\[
  \Phi =\sum_{i,j \geq 2}
   \langle v_i,\tilde v_j\rangle^2
   \int_0^\infty
   \bigl(
     e^{-2\widetilde{\lambda}_j t}
     +
     e^{-2\lambda_i t}
     -
     2\,e^{-(\lambda_i+\widetilde{\lambda}_j)t}
   \bigr)\,dt 
   \]
   \[
 =\sum_{i,j \geq 2}
   \langle v_i,\tilde v_j\rangle^2
   \Biggl(
     \frac{1}{2\widetilde{\lambda}_j}
     +
     \frac{1}{2\lambda_i}
     -
     \frac{2}{\lambda_i+\widetilde{\lambda}_j}
   \Biggr)
   \]

Simplifying yields: 

\begin{equation} 
    \Phi = \sum_{i,j \geq 2}
  \langle v_i,\tilde v_j\rangle^2
  \frac{(\lambda_i-\widetilde{\lambda}_j)^2}
       {2\,\lambda_i\,\widetilde{\lambda}_j\,(\lambda_i+\widetilde{\lambda}_j)}
\end{equation}

\section{Effective Resistance Sampling}
\label{sec:er_sampling}

\begin{figure}[h!]
    \centering
    \includegraphics[width=1\linewidth]{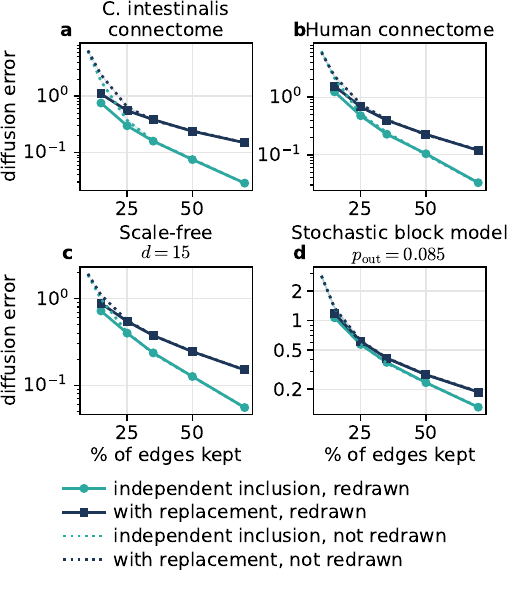}
    \caption{Effective Resistance sampled with independent inclusion  (teal) and i.i.d. with replacement, (dark blue). Solid lines: redrawn until connected, as in Figure \ref{fig:linear_diffusion_error}. Dotted lines: not redrawn, so some sparse graphs are disconnected. Linear diffusion trajectory error is measured as in Figure \ref{fig:linear_diffusion_error}, with the same initial conditions, draws and network realizations.}
    \label{fig:er_sampling}
\end{figure}

Figure~\ref{fig:er_sampling} compares the independent-inclusion sampler used throughout the paper with the original sampling with replacement algorithm presented in \cite{spielman_graph_2011}. At the same edge budget, independent inclusion gives lower diffusion error on every network, with the advantage growing with edge budget. With independent inclusion sampling, edges with the highest $w_eR_e$ are given $p_{ij} = 1$, and are kept with certainty at their original weight. Figure~\ref{fig:er_sampling} also shows that the redraw until connected variants of Effective Resistance sparsification yield similar diffusion error, except at very low edge budgets. 

\section{Optimization Details}
\label{sec:optimization_details}

\begin{algorithm}[H]
  \caption{Sparsification by minimizing an error $\mathcal{E}$}
  \label{alg:optimization}
  \begin{algorithmic}[1]
  \Require graph $G$ with weights $w$, the number of edges to keep $T$, and error $\mathcal{E}(\widetilde{G})$
  \Ensure sparse graph $\widetilde{G}$ with $T$ edges and weights $\tilde w$
  \State $\widetilde{G}\gets$ maximum-weight spanning tree of $G$, with $\tilde w_{ij} = w_{ij}$
  \While{$\widetilde{G}$ has fewer than $T$ edges}
    \parState{add the edge $(i,j) \notin \widetilde{G}$, with $\tilde w_{ij} = w_{ij}$, that most reduces $\mathcal{E}$, computing $\mathcal{E}$ exactly for the 5 edges whose addition is estimated, to first order, to lower it most}
  \EndWhile
  \parState{rescale all $\tilde w$ by a common factor, $(n-1)/\sum_{(i,j)\in \widetilde{G}} \tilde w_{ij} R_{ij}$, where $R_{ij}$ is the effective resistance of $(i,j)$ in $G$}
\For{each of two rounds}
    \For{up to 50 swaps}
      \parState{estimate, to first order, how much removing each edge in $\widetilde{G}$ or adding each edge not in $\widetilde{G}$ would change $\mathcal{E}$}
      \parState{pair the 20 removals and the 20 additions estimated to lower $\mathcal{E}$ most, and rank these $20 \times 20$ swaps by their estimated change in $\mathcal{E}$}
      \parState{going down the ranking, skip swaps that would disconnect $\widetilde{G}$, and compute $\mathcal{E}$ exactly for the first 5 others; if none lowers $\mathcal{E}$, continue until one does; stop if none does}
      \parState{for the swap that lowers $\mathcal{E}$ most, remove $(i,j)$ from $\widetilde{G}$ and add $(k,l)$ with $\tilde w_{kl} = (\tilde w_{ij}/w_{ij})\, w_{kl}$, keeping the removed edge's scale factor}
    \EndFor
    \parState{minimize $\mathcal{E}$ over the log-weights $\log \tilde w$ of the edges in $\widetilde{G}$ with L-BFGS-B (at most 60 iterations), keeping each $\tilde w_{ij}$ at most $20 \max w$}
  \EndFor
  \State \Return $\widetilde{G}$
  \end{algorithmic}
\end{algorithm}

\subsection{Uniform Spectral Error sparsification: details}
\label{section: USE details}

Algorithm \ref{alg:optimization} needs a smooth loss function, but $\varepsilon$ in (\ref{eq:seo_error}) is not always differentiable. We therefore replace the maximum by the $p$-norm of the same errors,
\begin{equation}
    \label{eq:p_norm}
      \varepsilon_p = \Big(\sum_{i=2}^{n} |\mu_i - 1|^p\Big)^{1/p},
\end{equation}
with $\mu_i$ the nonzero-mode eigenvalues of $L^{+/2}\tilde L L^{+/2}$. Eq. (\ref{eq:p_norm})
is smooth and approaches $\varepsilon$ as $p$ grows ($\varepsilon \le \varepsilon_p \le (n-1)^{1/p}\varepsilon$). We use continuation in $p$, where after the maximum spanning tree and fill stages (using $p = 8$), we first run the swaps and weight optimization steps at $p = 8$ (repeated twice), then $32$ then $128$, each run starting from the previous graph, which empirically works better than starting at large $p$. By the bound above, the final objective $\varepsilon_{128}$ is within a factor $(n-1)^{1/128}$ of $\varepsilon$, which for a graph of 100 nodes is within 4\%.

\section{Kuramoto integration and $K_{\mathrm{sync}}$}
\label{sec:K_sync}
All Kuramoto simulations are integrated on $t \in [0, 40]$ with an adaptive Runge--Kutta 4(5) integrator (relative tolerance $10^{-6}$, absolute tolerance $10^{-8}$), with the phases recorded at 400 evenly spaced times. A seed sets the natural frequencies, $\omega_i \sim \mathcal N(5, 0.5^2)$, and the initial phases, drawn uniformly on $[0, 2\pi)$, and at each coupling the full and sparse networks are integrated from the same seed. We measure coupling in units of $K_{\mathrm{sync}}$. The late-time order parameter is $r(t)$ averaged
over the last 30\% of the run, $t \in [28, 40]$, and $K_{\mathrm{sync}}$ is the coupling at which the full
network's $r$, averaged over 8 seeds, equals $0.5$, found by root finding. $K_{\mathrm{sync}}$ is computed
for each network (and each realization of a generated network), so that every network's synchronization
transition sits near $K/K_{\mathrm{sync}} = 1$.

\section{Synchronization Transition Curves}
\label{sec:synch_curves}
In Figures \ref{fig:phase_transition_curve} and \ref{fig:phase_transition_curves}, for each network and each fraction of edges kept, 10\%, 15\%, 25\%, 40\% and 60\%, we compute the late-time order parameter $r$ (Appendix~\ref{sec:K_sync}) at 36 couplings: $K/K_{\mathrm{sync}} = 0.05$, $0.1$,
$0.15$, from $0.2$ to $3$ in steps of $0.1$, and $4$, $6$, $8$, $10$, with $K_{\mathrm{sync}}$ of the full network. Generated networks use the first realization only (whereas some other analyses average over 10 realizations). Effective Resistance is drawn 20 times, each draw redrawn until connected, and each draw is run from 5 initial conditions (natural frequencies and initial phases, drawn as in
Appendix~\ref{sec:K_sync}), for 100 runs. DSE, USE and the full network are run from the same 100 initial conditions. Each curve is the mean of $r$ over these runs, and the curve RMSE is the root mean square, over the 36 couplings, of the difference between a sparse curve and the full network's curve. Effective
Resistance is left out where fewer than 1 in 1000 single draws are connected, and a sparsification level is left out where it keeps fewer edges than a spanning tree.

\begin{figure}
    \centering
    \includegraphics[width=0.9\linewidth]{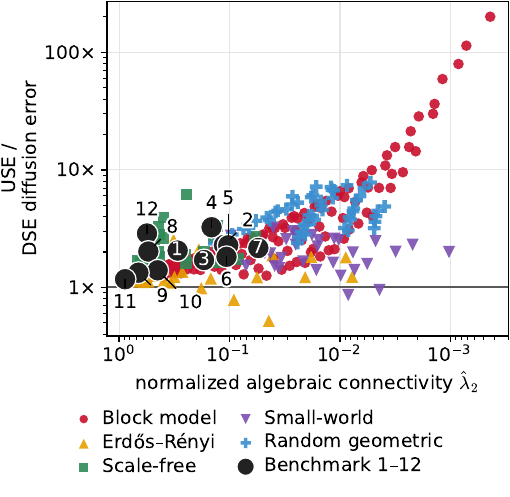}
    \caption{DSE's advantage over Uniform Spectral Error (USE) on linear diffusion trajectory error against bottleneck strength, keeping 25\% of each network's edges. The y-axis shows linear diffusion error of USE divided by that of DSE, so that values $>1$ mean that DSE performed better, and the x-axis shows the normalized algebraic connectivity $\hat\lambda_2$, where stronger bottlenecks are to the right. Each colored point is one parameter setting of a generator model, averaged over its realizations (40 for the stochastic block models, 10 for the others) and placed at the median $\hat\lambda_2$ of those realizations. Black circles are the twelve benchmark networks, numbered as in Appendix~\ref{sec:all_networks}. Linear diffusion error is measured as in Figure \ref{fig:linear_diffusion_error}.}
    \label{fig:DSE_advantage_use}
\end{figure}

\begin{figure} 
    \centering
    \includegraphics[width=0.9\linewidth]{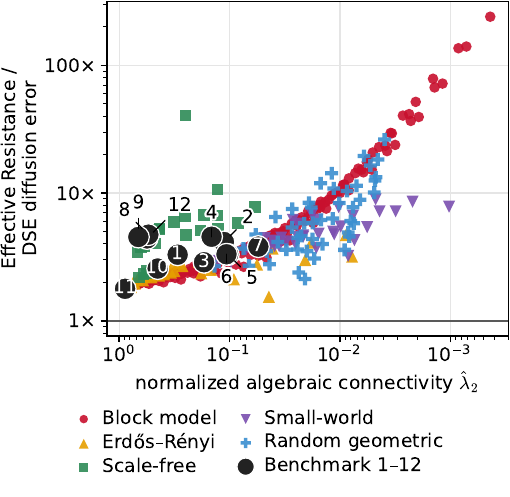}
    \caption{DSE's advantage over Effective Resistance on linear diffusion trajectory error against bottleneck strength. Details match those in Figure \ref{fig:DSE_advantage_use}.}
    \label{fig:DSE_advantage_er}
\end{figure}

\begin{figure*}
    \centering
    \includegraphics[width=1\linewidth]{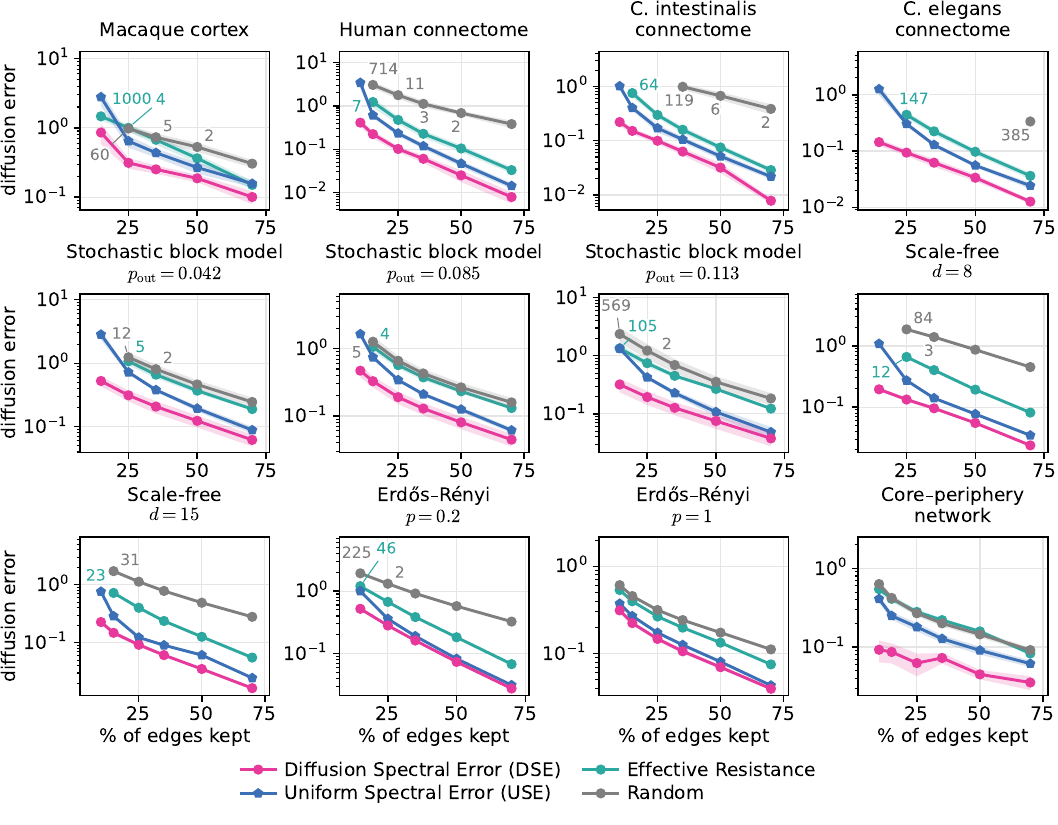}
    \caption{DSE outperforms other sparsifiers in preserving linear diffusion dynamics on all twelve networks. Linear diffusion trajectory error against the fraction of edges kept, as measured in Figure \ref{fig:linear_diffusion_error}: each point is the mean over 15 random initial conditions, $x_0 \sim \mathcal N(0, I)$. For each initial condition, errors are first averaged over the 10 Effective Resistance or Random network draws and, for generated networks, over the 10 network realizations. Effective Resistance and Random are redrawn until connected, and printed numbers next to individual points give the mean number of draws needed to yield a connected network.}
    \label{fig:diffusion_error_all}
\end{figure*}

\begin{figure*}
    \centering
    \includegraphics[width=1\linewidth]{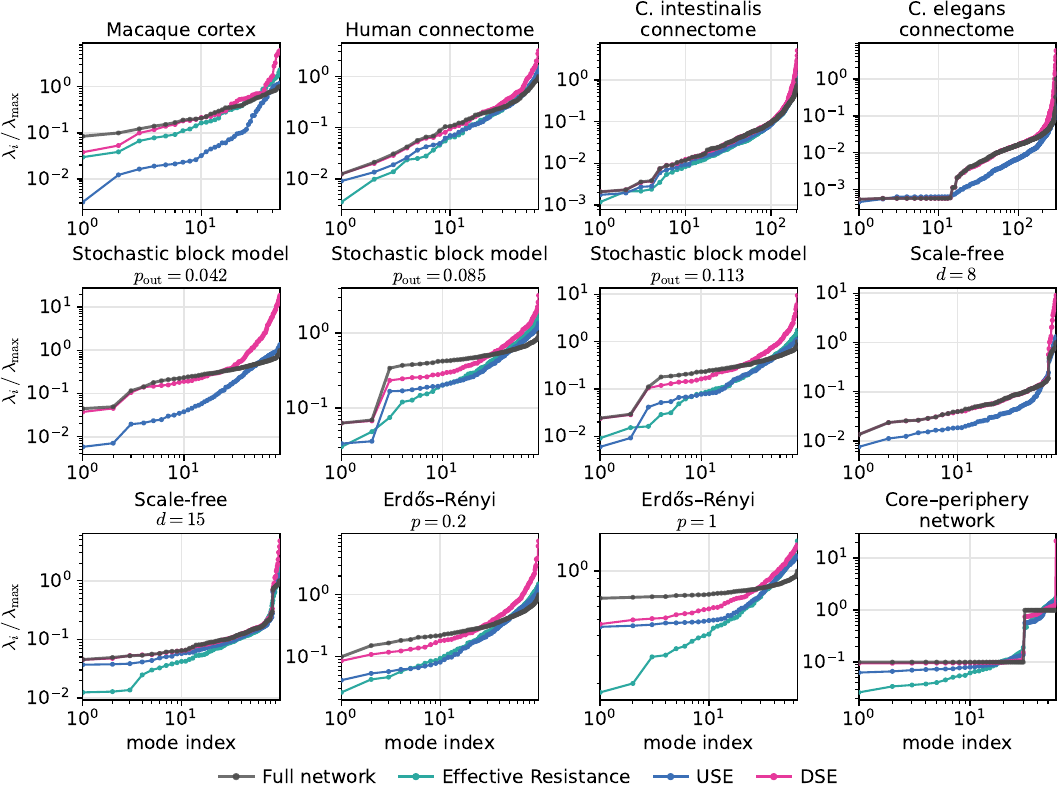}
    \caption{Laplacian eigenvalues of all twelve networks and three sparsifications keeping 15\% of their edges, as in Figure \ref{fig:connectome66_eigenvalues}. The eigenvalues are sorted from least to greatest, and normalized by the largest eigenvalue of the full graph. Across networks (except Macaque cortex and Erdős–Rényi), DSE closely matches the small eigenvalues (slow modes) and overshoots the large eigenvalues (fast modes), while Uniform Spectral Error sparsification and Effective Resistance frequently underweight slow modes. Effective Resistance is redrawn until connected, and is left out of a panel if fewer than 1 in 1000 draws are connected.}
\label{fig:eigenvalues_all_networks}
\end{figure*}

\begin{figure*}
    \centering
    \includegraphics[width=1\linewidth]{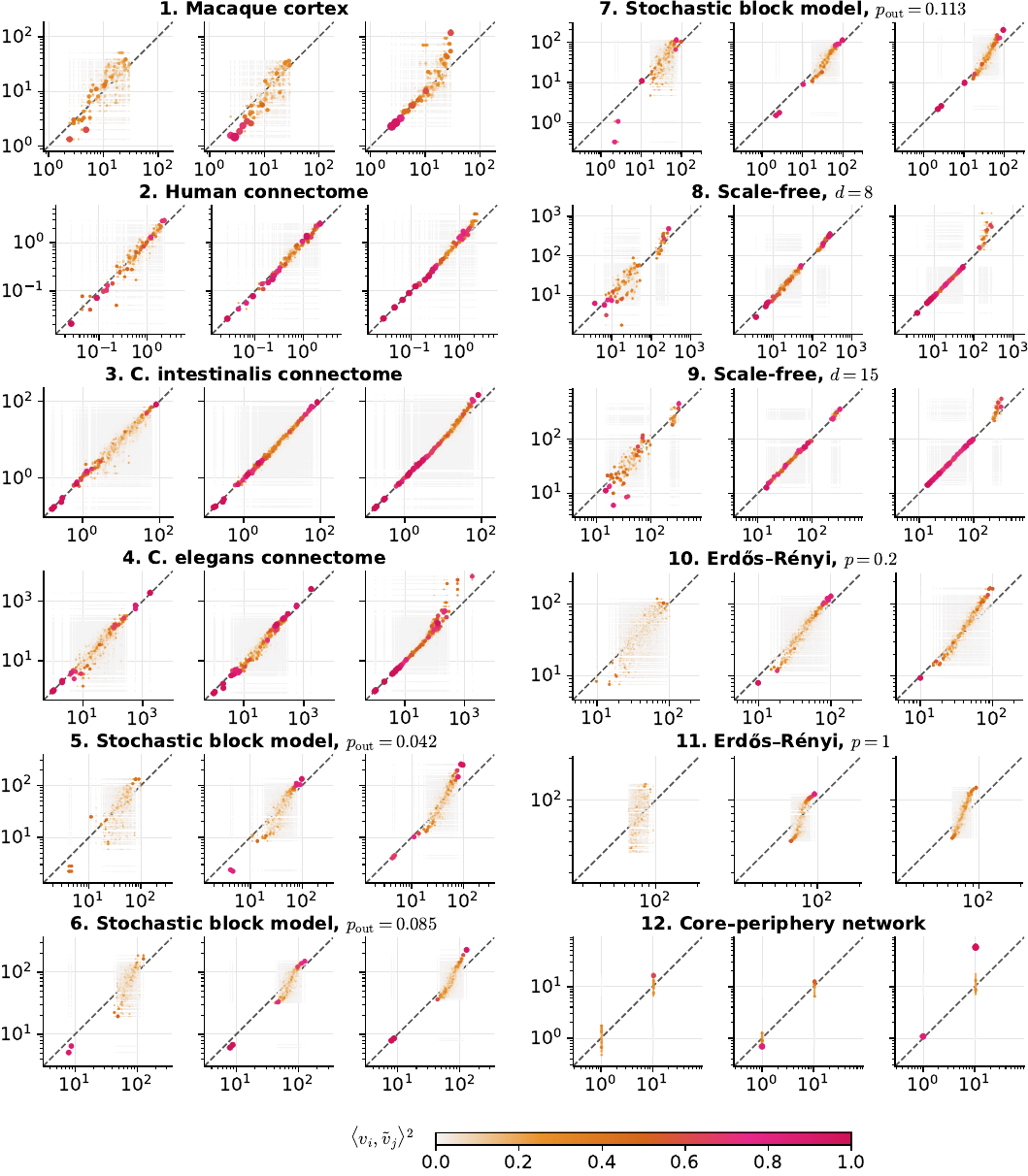}
    \caption{Eigenvector overlap between all twelve networks and three sparsifiers keeping 25\% of their edges. Each block shows one network, with panels from left to right for Effective Resistance, USE, and DSE. As in Figure \ref{fig:eigenoverlap}, at every pair of modes between the original and sparse networks, ($\lambda_i$, $\widetilde{\lambda}_j$), we place a dot where the color and size show $\langle v_i, \tilde v_j\rangle^2$, which shows how aligned the corresponding eigenvectors are. A perfect sparsifier would only have colored dots on the diagonal. Zero modes are left out.}
    \label{fig:overlap_all}
\end{figure*}

\begin{figure*}
    \centering
    \includegraphics[width=1\linewidth]{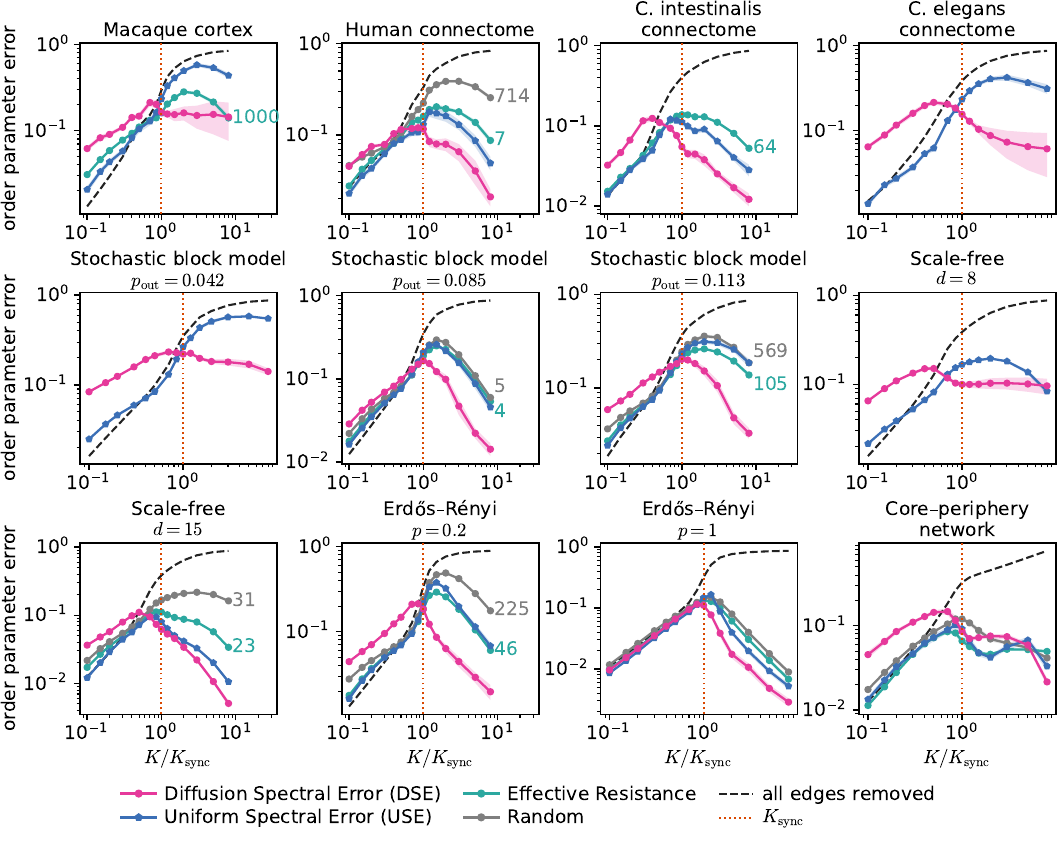}
    \caption{Kuramoto order-parameter error by coupling strength at 15\% of edges kept, for all twelve networks. Order-parameter error is measured and averaged as in Figure \ref{fig:order_parameter_error}. The x-axis sweeps over global coupling $K$ in units of each network's $K_{\mathrm{sync}}$, the coupling at which the full network's late-time order parameter reaches $r = 0.5$. The shading shows $\pm 1$ standard error of the mean over seeds. Sampled networks are left out of a panel entirely if fewer than 1 in 1000 draws are connected. The number beside the right end of the Random and Effective Resistance curves gives the mean number of draws needed to yield a connected graph. The dashed black line shows the order-parameter error with $K = 0$.}
    \label{fig:order_parameter_all}
\end{figure*}

\begin{figure*}
    \centering
    \includegraphics[width=1\linewidth]{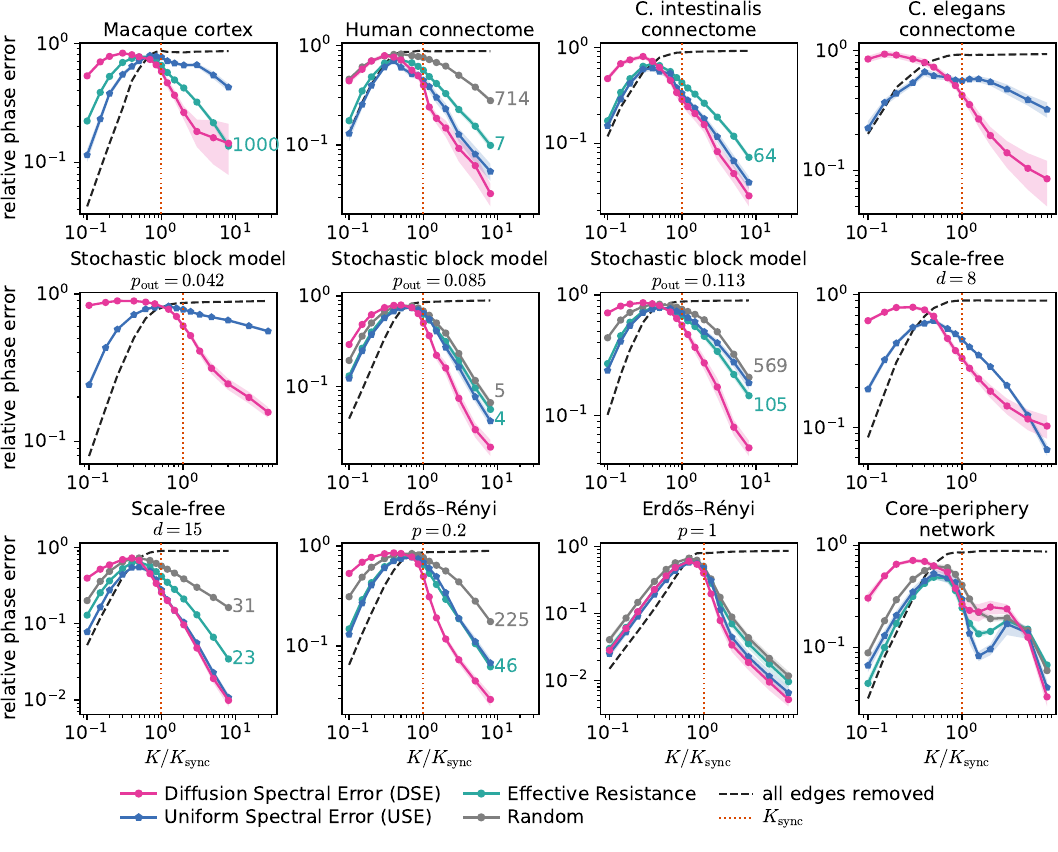}
    \caption{Kuramoto relative phase error by coupling strength at 15\% of edges kept, for all twelve networks. Relative phase error is the trajectory error of Eq. \eqref{eq:kuramoto_error} measured after subtracting each run's mean phase from every oscillator, so that a uniform rotation of all phases does not count as error, as in Figure \ref{fig:kuramoto_trajectory}b. The x-axis, shading, and printed numbers are as in Figure \ref{fig:order_parameter_all}. The dashed black line shows the relative phase error with $K = 0$.}
    \label{fig:relative_phase_all}
\end{figure*}

\begin{figure*}
    \centering
    \includegraphics[width=1\linewidth]{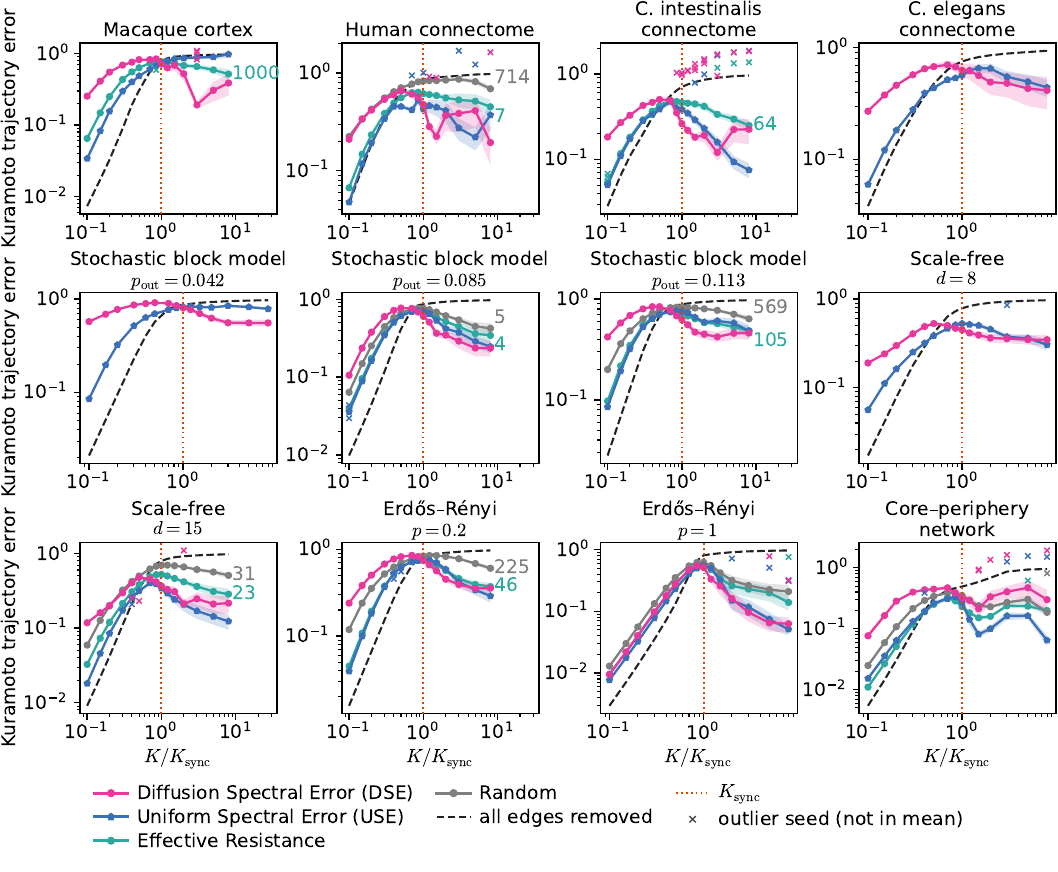}
    \caption{Kuramoto trajectory error, Eq. \eqref{eq:kuramoto_error}, by coupling strength at 15\% of edges kept, for all twelve networks, as in Figure \ref{fig:kuramoto_trajectory}a. The x-axis, shading, and printed numbers are as in Figure \ref{fig:order_parameter_all}. Crosses mark outlier seeds (beyond three interquartile ranges from the quartiles), which are left out of the mean. The dashed black line shows the trajectory error with $K = 0$.}
    \label{fig:kuramoto_trajectory_all}
\end{figure*}

\begin{figure*} 
    \centering
    \includegraphics[width=0.8\linewidth]{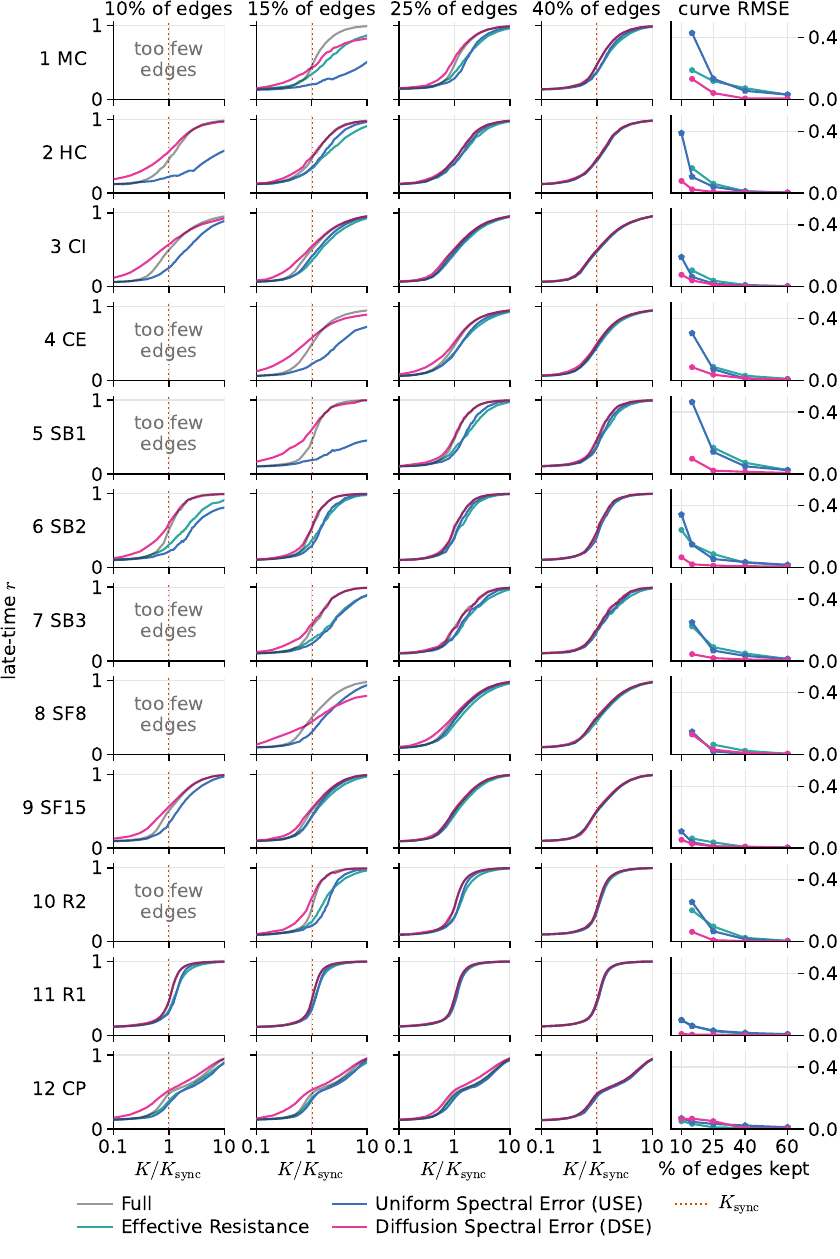}
    \caption{DSE best preserves the synchronization transition curve on all twelve networks except Core-periphery. Each row is one network, numbered as in Appendix~\ref{sec:all_networks}: MC macaque cortex, HC human connectome, CI \textit{C.~intestinalis}, CE \textit{C.~elegans}, SB1--SB3 stochastic block models 5--7, SF8 and SF15 scale-free networks with $d =8$ and $15$, R2 and R1 Erd\H{o}s--R\'enyi networks with $p = 0.2$ and $p = 1$, and CP core--periphery. The first four columns show the late-time order parameter $r$ against $K/K_{\mathrm{sync}}$ at 10\%, 15\%, 25\% and 40\% of edges kept, and the last column shows the root mean squared difference between each sparse curve and the full one, against the fraction of edges kept (10--60\%). Effective Resistance draws are redrawn until connected and left out where fewer than 1 in 1000 single draws connect.}
    \label{fig:phase_transition_curves}
\end{figure*}

\clearpage
\section{Networks}

\label{sec:all_networks}

\begin{figure*}
    \centering
    \includegraphics[width=1\linewidth]{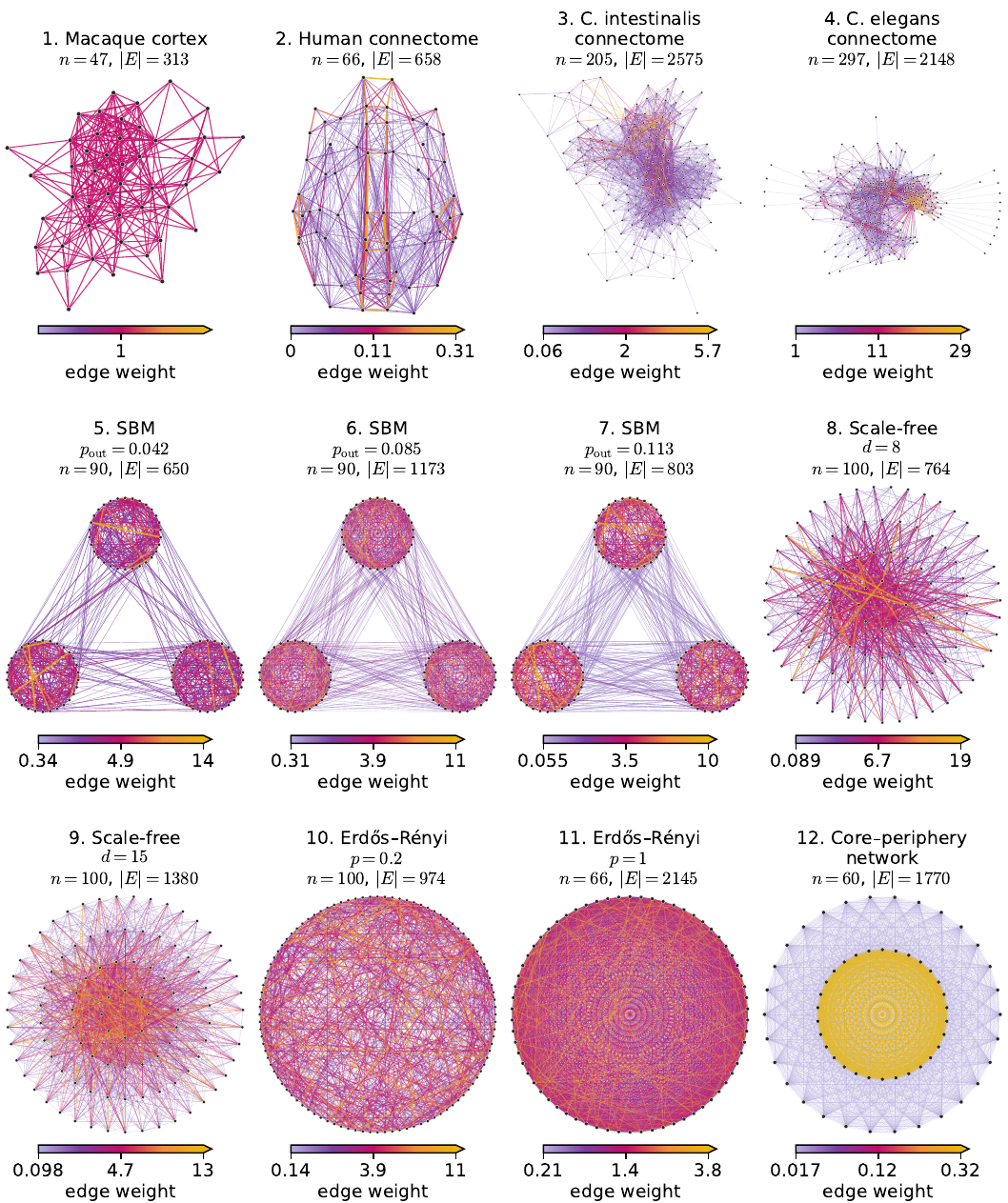}
    \caption{The twelve networks used in this paper, numbered as in the list above. Line width and color represent edge weight, with the color scale beneath each network.}
    \label{fig:networks_all}
\end{figure*}

The four empirical networks are taken from public repositories as described below. Every network is undirected, with positive weights and no self-loops.

\begin{enumerate}
\item \textbf{Macaque cortex ($n = 47$).}
Netzschleuder \cite{peixoto_netzschleuder_2020} entry \texttt{macaque\_neural}. The binary directed CoCoMac \cite{kotter_online_2004} matrix of \citet{honey_network_2007}: 47 areas on the \citet{felleman_distributed_1991} map and 505 directed links. An undirected edge of weight 1 is kept when at least one direction is present, leaving 313 edges.

\item \textbf{Human connectome ($n = 66$).}
\texttt{weights.txt} from \texttt{connectivity\_66.zip} in The Virtual Brain data package
\cite{sanz_leon_virtual_2013}. The 66 labels are the cortical regions of
\citet{hagmann_mapping_2008}, and the entries are connection strengths in arbitrary units. The
stored matrix $W$ is replaced by $(W+W^{\mathsf T})/2$ with the diagonal set to zero, leaving 658
edges.

\item \textbf{\textit{C.~intestinalis} connectome ($n = 205$).}
Netzschleuder entry \texttt{cintestinalis}, from the larval connectome of
\citet{ryan_cns_2016}. The file has 205 cells and 2903 directed edges. A directed weight is the
cumulative depth of the presynaptic contacts, in $\mu$m. The undirected weight is the sum of the
two directions, and the diagonal is set to zero, leaving 2575 edges.

\item \textbf{\textit{C.~elegans} connectome ($n = 297$).}
Netzschleuder entry \texttt{celegansneural}: the compilation by \citet{watts_collective_1998} of
the data of \citet{white_structure_1986}. It has 2359 directed edges. The weights are the
integers given by Watts and Strogatz, and the undirected weight is the sum of the two directions,
leaving 2148 edges.

\item \textbf{Stochastic block model ($n = 90$, $p_{\mathrm{out}} = 0.042$).}
A stochastic block model \cite{holland_stochastic_1983} with three blocks of 30 and
$p_{\mathrm{in}} = 0.35$. The value $p_{\mathrm{out}} = 0.0423$ is chosen so the expected fraction
of a node's degree that leaves its block is $f_{out} = 0.2$. Edges inside a block are
$\mathrm{LogNormal}(1, 0.5)$; edges between blocks are $\mathrm{LogNormal}(0, 0.5)$. Each block is
completed by a random tree, and consecutive blocks are joined by one bridge, using the same weight
law as that edge type.

\item \textbf{Stochastic block model ($n = 90$, $p_{\mathrm{out}} = 0.085$).}
Three blocks of 30 with denser blocks, $p_{\mathrm{in}} = 0.7$. The value
$p_{\mathrm{out}} = 0.0846$ is chosen so that again $f_{\mathrm{out}} = 0.2$. Weights are as in 5: within-block
$\mathrm{LogNormal}(1, 0.5)$ and between-block $\mathrm{LogNormal}(0, 0.5)$. The same tree and
bridge repair is applied. This is the stochastic block model shown in the main text.

\item \textbf{Stochastic block model ($n = 90$, $p_{\mathrm{out}} = 0.113$).}
Three blocks of 30 with $p_{\mathrm{in}} = 0.35$, as in network 5. The value $p_{\mathrm{out}} = 0.1128$ is
chosen so $f_{\mathrm{out}} = 0.4$. Within-block weights are $\mathrm{LogNormal}(1, 0.5)$; between-block weights
are $\mathrm{LogNormal}(-1.5, 0.5)$. The same tree and bridge repair is applied.

\item \textbf{Scale-free ($n = 100$, $d =8$).}
Preferential attachment \cite{barabasi_emergence_1999}. Start from a clique of 8 nodes. Each new
node adds 8 edges to distinct earlier nodes, with probability proportional to
$\mathrm{degree}+1$. An edge that touches a node at or above the 85th percentile of degree is
$\mathrm{LogNormal}(1.5, 0.5)$; every other edge is $\mathrm{LogNormal}(-1, 0.5)$.

\item \textbf{Scale-free ($n = 100$, $d =15$).}
Same construction as 8 with $d =15$. This is the scale-free network shown in the main text.

\item \textbf{Erd\H{o}s--R\'enyi ($n = 100$, $p = 0.2$).}
An Erd\H{o}s--R\'enyi graph \cite{erdos_random_1959}: edges are present independently with
probability $p = 20/99$, and the graph is redrawn until it is connected. Each edge weight is
$\mathrm{LogNormal}(\rho, 0.5)$ with $\rho$ drawn uniformly from $\{1.2,-0.8\}$.

\item \textbf{Erd\H{o}s--R\'enyi ($n = 66$, $p = 1$).}
The complete graph on 66 nodes. The 2145 undirected edges are i.i.d.\
$\mathrm{LogNormal}(0, 0.5)$.

\item \textbf{Core--periphery network ($n = 60$).}
A network with a designed Laplacian spectrum, following \citet{forrow_functional_2018}. Two
groups of 30: edges inside the core (group 1) have weight $19/60$; every other pair has weight
$1/60$. The Laplacian spectrum is $\{0,\, 1\ (\times 30),\, 10\ (\times 29)\}$.
\end{enumerate}

\section*{Code Availability}
An implementation of DSE and USE sparsification and of effective resistance and random sampling, along with the network generators and empirical network data used in this paper, is available at \url{https://github.com/emmiefitzgibbon/dynamics-preserving-graph-sparsification}.

\end{document}